\documentclass[twocolumn]{aastex63}

\usepackage{graphicx,textcomp,fancyhdr,hyperref,xcolor,fontawesome}
\definecolor{linkcolor}{rgb}{0.1216,0.4667,0.7059}
\usepackage[caption=false]{subfig}
\usepackage{float}
\usepackage{footnote}

\shorttitle{Habitable Zone Catalog}
\shortauthors{Michelle L. Hill et al.}

\begin{document}

\title{A Catalog of Habitable Zone Exoplanets}

\author[0000-0002-0139-4756]{Michelle L. Hill}
\affiliation{Department of Earth and Planetary Sciences, University of California Riverside, 900 University Ave, Riverside, CA 92521, USA}
\email{mhill012@ucr.edu}

\author[0000-0002-4420-0560]{Kimberly Bott}
\affiliation{Department of Earth and Planetary Sciences, University of California Riverside, 900 University Ave, Riverside, CA 92521, USA}  

\author[0000-0002-4297-5506]{Paul A. Dalba}
\altaffiliation{NSF Astronomy and Astrophysics Postdoctoral Fellow}
\affiliation{Department of Astronomy and Astrophysics, University of California, Santa Cruz, CA 95064, USA}
\affiliation{Department of Earth and Planetary Sciences, University of California Riverside, 900 University Ave, Riverside, CA 92521, USA}

\author[0000-0002-3551-279X]{Tara Fetherolf}
\altaffiliation{Chancellor's Postdoctoral Fellow}
\affiliation{Department of Earth and Planetary Sciences, University of California Riverside, 900 University Ave, Riverside, CA 92521, USA}

\author[0000-0002-7084-0529]{Stephen R. Kane}
\affiliation{Department of Earth and Planetary Sciences, University of California Riverside, 900 University Ave, Riverside, CA 92521, USA}

\author[0000-0002-5893-2471]{Ravi Kopparapu}
\affiliation{NASA NExSS Virtual Planetary Laboratory, Seattle, WA USA}
\affiliation{Blue Marble Space Institute of Science, Seattle, WA USA}
\affiliation{NASA Goddard Space Flight Center, Greenbelt, MD USA}

\author[0000-0002-4860-7667]{Zhexing Li}
\affiliation{Department of Earth and Planetary Sciences, University of California Riverside, 900 University Ave, Riverside, CA 92521, USA}

\author[0000-0002-7084-0529]{Colby Ostberg}
\affiliation{Department of Earth and Planetary Sciences, University of California Riverside, 900 University Ave, Riverside, CA 92521, USA}


\begin{abstract}

The search for habitable planets has revealed many planets that can vary greatly from an Earth analog environment. These include highly eccentric orbits, giant planets, different bulk densities, relatively active stars, and evolved stars. This work catalogs all planets found to reside in the HZ and provides HZ boundaries, orbit characterization, and the potential for spectroscopic follow-up observations. Demographics of the HZ planets are compared with a full catalog of exoplanets. 
Extreme planets within the HZ are highlighted, and how their unique properties may affect their potential habitability.
Kepler-296~f is the most eccentric $\leq~2~R_\oplus$ planet that spends 100\% of its orbit in the HZ. HD~106270~b and HD~38529~c are the most massive planets ($\leq~13~M_J$) that orbit within the HZ, and are ideal targets for determining the properties of potential hosts of HZ exomoons. These planets, along with the others highlighted, will serve as special edge-cases to the Earth-based scenario and observations of these targets will help test the resilience of habitability outside the standard model. The most promising observational HZ target that is known to transit is GJ~414~A~b. Of the transiting, $\leq~2~R\oplus$ HZ planets, LHS~1140~b, TRAPPIST-1~d and K2-3~d are the most favorable. Of the non-transiting HZ planets HD~102365~b and 55~Cnc~f are the most promising, and the best non-transiting candidates that are $\leq~2~R_\oplus$ are GJ~667~C~c, Wolf~1061~c, Ross~508~b, Teegarden's~Star~b, and Proxima~Cen~b.

\end{abstract}



\section{Introduction}
\label{intro}

\begin{figure}
  \begin{center}
    \includegraphics[width=0.5\textwidth]{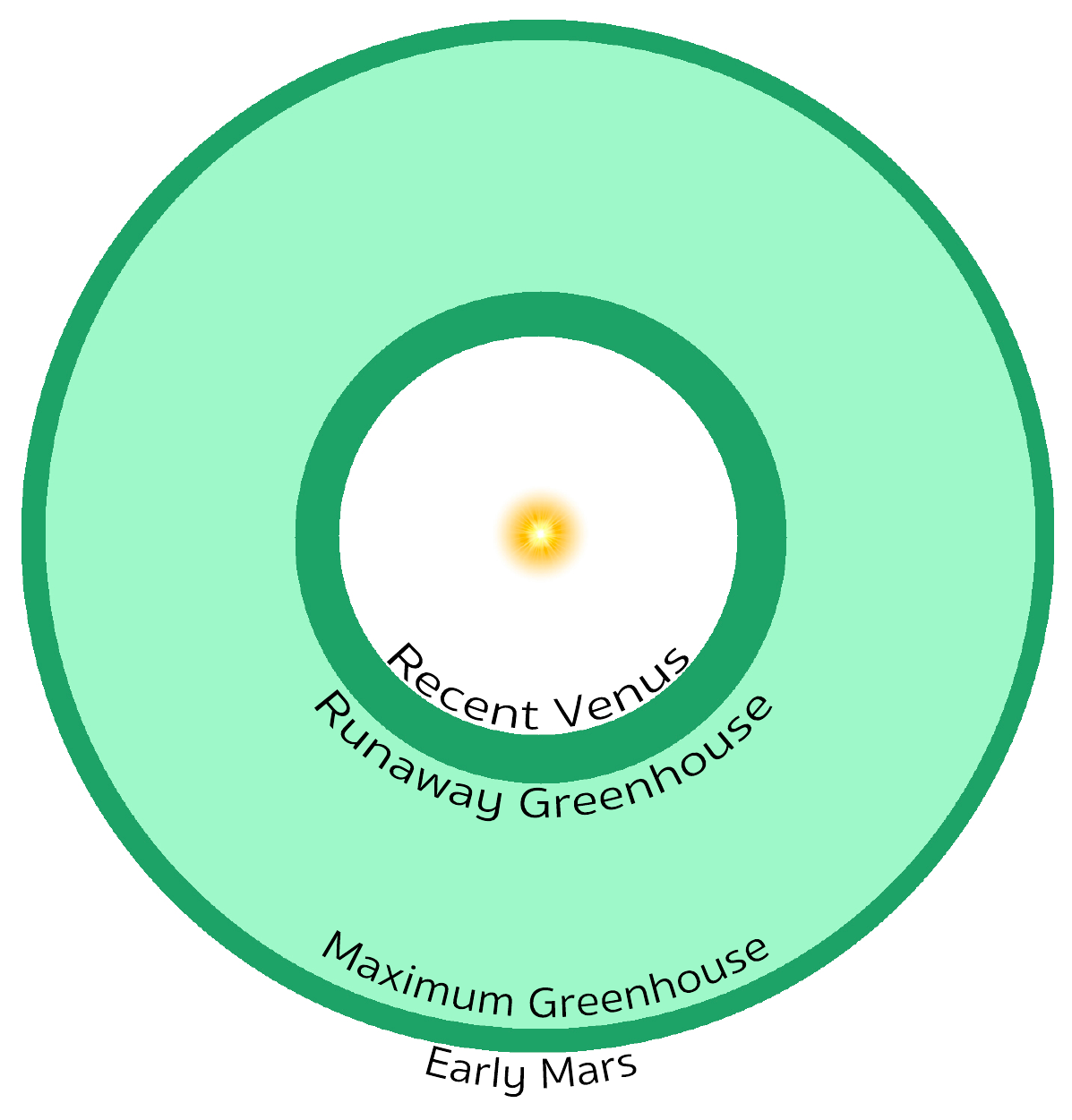}
  \end{center}
  \caption[Habitable Zone Boundaries]{Depiction of both the conservative (light green) and optimistic (combined light and dark green) HZ regions, with boundary labels defined by \citet{kopparapu2013a}.}
  \label{fig:HZbound}
\end{figure}

Exoplanet discoveries over the past several decades have revealed a vast diversity of planetary architectures \citep{ford2014,winn2015}, and shown that terrestrial planets are far more common than their giant planet counterparts \citep{borucki2016}.
In these ongoing exoplanet searches, discovering those planets that may harbor life has been a primary objective for the astrobiology community \citep{fujii2018,schwiererman2018,glaser2020,lisse2020}. A potential pathway toward the identification of such worlds is to constrain the stellar and planetary parameter space that may allow for the presence of surface liquid water. Such is the premise of the habitable zone (HZ), defined as the region around a star where water can exist in a liquid state on the surface of a planet with sufficient atmospheric pressure \citep{kasting1993a,kopparapu2013a,kopparapu2014,kane2016c}. The HZ broadly consists of two main regions; the conservative habitable zone (CHZ) and the optimistic habitable zone (OHZ), shown in Figure~\ref{fig:HZbound} as the light green and dark green regions, respectively. The CHZ inner boundary is the runaway greenhouse limit, during which water loss can occur through photodissociation of water molecules in the upper atmosphere. The CHZ outer boundary is the maximum greenhouse, where the planetary temperature conditions allow condensation of substantial atmospheric $CO_2$ on the surface \citep{kopparapu2013a}. The OHZ inner boundary, the Recent Venus limit, is based on the empirical observation that the surface of Venus has been dry for at least a billion years, but may have had conditions suitable for surface liquid water prior \citep{kane2014e,way2016}. The outer edge of the OHZ, the Early Mars limit, is based on evidence that Mars appears to have harbored surface liquid water $\sim$3.8~Gya \citep{kopparapu2013a}. 

The HZ has undergone numerous revisions to incorporate scenarios quite different to Earth, such as tidal locking \citep{yang2014b}, desert environments \citep{abe2011b}, and substantial methane greenhouse warming \citep{ramirez2018c}. Although these revisions often extend the possible surface liquid water environments, it has also been suggested that the HZ may not be restrictive enough for supporting complex life \citep{Schwieterman2019}. Even so, the HZ has been a critical component of calculating the occurrence rate of terrestrial planets potentially amenable to temperate surface conditions \citep{dressing2013,bryson2021} and the demographics of the HZ planet population \citep{adams2016b}. These methodologies have been extended to consider the occurrence of giant planets in the HZ that may harbor temperate moons \citep{heller2012,hill2018} and the prevalence of potential Venus analogs \citep{kane2014e}. Dynamical simulations have been utilized to determine regions of orbital stability within the HZ that could harbor further terrestrial planets \citep{kopparapu2010,kane2020b}. Of the thousands of known exoplanets, hundreds reside within the HZ \citep{kane2012a}, and the extent of the HZ has been calculated for many nearby stars using available stellar data \citep{chandler2016,kopparapu2018}.
The tracking and cataloging of HZ planets is an important task for the purposes of designing efficient follow-up strategies to further characterize their planetary properties \citep{kane2016c,kopparapu2018}. This need is particularly heightened in the era of the Transiting Exoplanet Survey Satellite \citep[TESS;][]{ricker2015}, whose exoplanet discoveries around bright host stars will be particularly amenable to near-term atmospheric characterization \citep{kempton2018}. On longer time scales, mission concepts, such as LUVOIR \citep{reportluvoir}, the Habitable Exoplanet Observatory \citep[HabEx;][]{reporthabex}, and the recommendations of the Astro2020 report \citep{Astro2020}, will provide further opportunities to understand the current state and evolution of habitable surface environments. As thousands of exoplanets have already been discovered, and both current and future exoplanet missions will only increase that number, the HZ is still an essential tool to refine the list of potentially habitable planets for target selection in future observational campaigns.

In this paper, we present a catalog of HZ planets that includes all known exoplanetary systems, irrespective of detection method, for which their properties allow the determination of their HZ status. This catalog updates previous HZ catalogs and extends the catalog to include all HZ planets, not just those from a specific mission (i.e. Kepler \citep{gaidos2013b, hill2018}). 
The focus of this HZ catalog is the demographics of the known HZ exoplanet population and the prospects for follow-up observations that will best serve the community in the fulfillment of the Astro2020 report recommendations.
In Section~\ref{meth}, we describe the data extraction, criteria for inclusion, and the calculations for the catalog assembly. Section~\ref{cat} presents the main catalog data for the exoplanets that meet our selection criteria, including measured and calculated properties for the stars and planets. In Section~\ref{discussion}, we provide a discussion of the catalog demographics, including analysis of outliers and the primary targets for follow-up opportunities. We outline the overall features of the catalog, proposed expansion and exploitation opportunities, and further concluding remarks in Section~\ref{conclusions}.


\section{Data Extraction and Calculations}
\label{meth}



The full data set of known exoplanets was downloaded from the NASA Exoplanet Archive \citep{akeson2013,ps} using the Application Programming Interface (API). The default values for each planet were downloaded and are used in the calculations and discussion that follows. More about the use of only the default values can be found in Section~\ref{discussion}. Planets with no stellar effective temperature ($T_\mathrm{eff}$) were removed from the list. For planets that were missing stellar luminosity ($L_\odot$), luminosity was calculated using the Stefan-Boltzmann law. When unavailable, the stellar radius ($R_{\odot}$) was calculated from the stellar mass ($M_\odot$) and surface gravity ($\log~g$). Stellar luminosities, and therefore HZ boundaries, can be sensitive to the uncertainties associated with their calculation \citep{kane2014a,chandler2016,kane2018a}.
In cases where luminosity was missing and could not be calculated, the planet was removed from the sample. The stellar mass and planet orbital period ($P$) were used to calculate the semi-major axis ($a$) of planets if it was missing from the NEA. If the semi-major axis data was missing and it was not able to be calculated, the planet was removed from the sample. 

HZ boundaries were calculated via the method described by \citet{kopparapu2013a, kopparapu2014}. Four HZ boundaries were calculated for each planet and are presented in Table \ref{tab:HZpl}: Recent Venus (R1), Runaway Greenhouse (R2), Maximum Greenhouse (R3), and Early Mars (R4). The percentage of time each planet spent in the CHZ and OHZ was calculated by solving Kepler's equation for each planets orbit. Planets were included in the HZ table if they spent any amount of time in the OHZ. We chose to include all planets that spent time in the HZ, no matter how briefly, to account for the uncertainty in HZ boundaries and orbital parameters. 
Table~\ref{tab:HZpl} includes a column that indicates how much of a planet's orbit is spent within the CHZ and OHZ. The list also includes planets of all sizes, regardless of whether they are expected to be in the terrestrial or gaseous regime. This is to include any planets that may be host to terrestrial exomoons, which are also potentially habitable worlds. 

Only 15 planets on the list had both a mass and radius measurement, the remaining planets having only one or the other measured, and so the mass-radius relationship from \citet{chen2017} was used to determine these values for planets that were missing either mass or radius information. We used the probabilistic modeling tool \texttt{forecaster} to estimate planet mass or radius based on the mass-radius relation from \citet{chen2017} that spans from dwarf planets
to late-type stars. We chose to use the \citet{chen2017} relation for consistency across our sample of planets, although we note that there are other mass-radius relations that may be well suited to specific planet regimes (such as \citet{thorngren2019} for cool giant planets, and \citet{Wolfgang2016, weiss2014} for planets $<4$~$R_\oplus$). We recommend further investigation into these other mass-radius relations for any individual planets the reader is interested in that is missing mass or radius measurement information, particularly for the planets with measured radii $>~9~R_\oplus$ for which mass measurements were estimated, as there is a large degeneracy in the mass estimations from \citet{chen2017} in this region. The mass and radius values that were calculated using the mass-radius relationship are included in Table \ref{tab:HZpl} and are distinguished by \textit{italic font} and are missing uncertainty values. 

Planet equilibrium temperature ($T_\mathrm{eq}$) was calculated from luminosity (assuming an albedo of 0) and subsequently used to calculate the transmission spectroscopy metric (TSM) value of each planet \citep{kempton2018}. The TSM is proportional to the expected transmission spectroscopy signal-to-noise, based on the strength of spectral features, brightness of the host star, and mass and radius of the planet assuming cloud-free atmospheres. The method for calculating the TSM is only applicable for planets $<10$~$R_\oplus$, therefore TSM is not included for planets with larger radii in Table~\ref{tab:HZpl}. We calculate TSM for all planets, not just those known to transit, to account for planets that may yet be found to transit and to allow direct comparison of atmospheric characterization in a broader demographic context. TSM values shown in Table~\ref{tab:HZpl} for planets where planet radii is calculated from \citet{chen2017} are provided for future use in the case where the planet is later found to transit and should not be taken as an indication that the planet transits.

To account for any planets whose eccentricity values were set to zero as a default value, rather than measured, eccentricity values were filtered for the histogram and scatter plots to only include planets with eccentricity uncertainty measurements. Additionally, the planets that have eccentricity fixed to zero are listed as 0 in Table~\ref{tab:HZpl}, whereas eccentricities that were measured as 0 are listed as 0.00. Note that those eccentricities that are fixed to zero means that they may be considered lower limits on those values.


\section{Catalog of HZ Exoplanets}
\label{cat}


Here we present the catalog of HZ exoplanets. Each planet listed in these tables passes through the OHZ at some point in their orbit. No cutoff has been made for the amount of time the planet must spend in the HZ. As planets that orbit outside the HZ temporarily are expected to be able to maintain or regain habitable periods \citep{williams2002, kane2012e, way2017a, palubski2020a, Kane2021}, planets that spend any time in the HZ (from here referred to as $>0\%$HZ) are included in the catalog. 

Table~\ref{tab:HZpl} includes the planet parameters for the HZ planets, and Table~\ref{tab:HZstar} includes the stellar parameters. The two tables are available combined as one master table in machine-readable format and can be filtered by the ``\% in HZ" columns as needed. Table~\ref{tab:HZpl} is ordered by the \% time the planet spends in the CHZ followed by time in the OHZ then followed by the planet name. Table \ref{tab:HZstar} is ordered by star name. 

Planet mass ($M_P$) and radius ($R_P$) are included in Table~\ref{tab:HZpl}. Measured values of each include uncertainty values in the table. Mass and radius values that have been calculated using the mass-radius relationship from \citet{chen2017} are distinguished with both \textit{italic font} and are missing uncertainty values.

Columns R1, R2, R3, and R4 list the Recent Venus, Runaway Greenhouse, Maximum Greenhouse, and Early Mars HZ boundaries respectively. Planetary orbital period ($P$), semi-major axis ($a$),  incident flux (flux) and eccentricity ($e$) are also included in the Table~\ref{tab:HZpl}. 

Planets whose TSM values are missing lack $T_\mathrm{eq}$ information, or the means to calculate it, or their parameter values were outside the limits set in the paper (i.e. $\geq 10$~$R_\oplus$). 

Within Tables~\ref{tab:HZpl} and \ref{tab:HZstar} and denoted by an exclamation point ($!$) next to the planet name are six planets that are marked as controversial in the NEA at the time of printing. These are HD~40307~g, KIC~5951458 b, GJ~667~C~e, GJ~667~C~f, Kepler-452~b, and Kepler-186~f \citep{diaz2016b, dalba2020b,robertson2014,mullally2018,burke2019}. As these planets are still listed as confirmed in the NEA we have elected to retain these planets in our sample. However, we recommend further investigation into their validity if any of these planets are of interest for follow-up. Further discussion as to why these planets are listed as controversial is included in Section~\ref{controversy}.

There are five circumbinary planets (CBPs) that have orbits that spend time the HZ. These are listed separately in Table~\ref{tab:cb}. The table includes planet mass ($M_p$), radius ($R_p$), orbital period ($P$), semi-major axis ($a$), eccentricity ($e$), stellar mass of primary ($M_{\star,\mathrm{A}}$), stellar radius of primary ($R_{\star,\mathrm{A}}$), effective temperature of primary ($T_{\mathrm{eff},\mathrm{A}}$), $\log~g$ of primary, luminosity of primary ($L_{\star,\mathrm{A}}$), $J$ magnitude ($J$), stellar mass of secondary ($M_{\star,\mathrm{B}}$), stellar radius of secondary ($R_{\star,\mathrm{B}}$). Further discussion of these systems is in Section~\ref{cb}.
 
\begin{figure*}
\centering 
\subfloat{%
  \includegraphics[width=1\columnwidth, height=0.8\columnwidth]{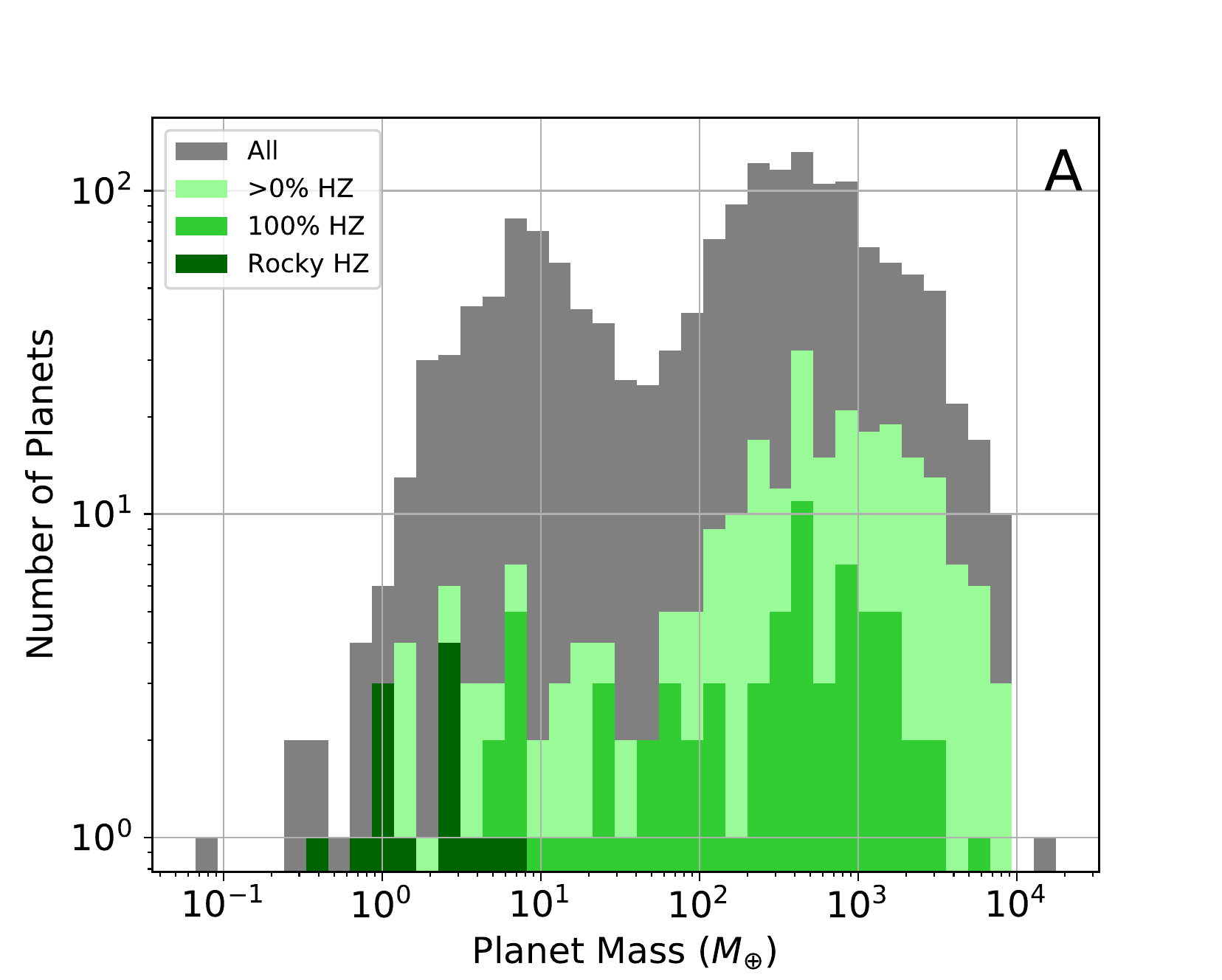}%
}
\subfloat{%
  \includegraphics[width=1\columnwidth, height=0.8\columnwidth]{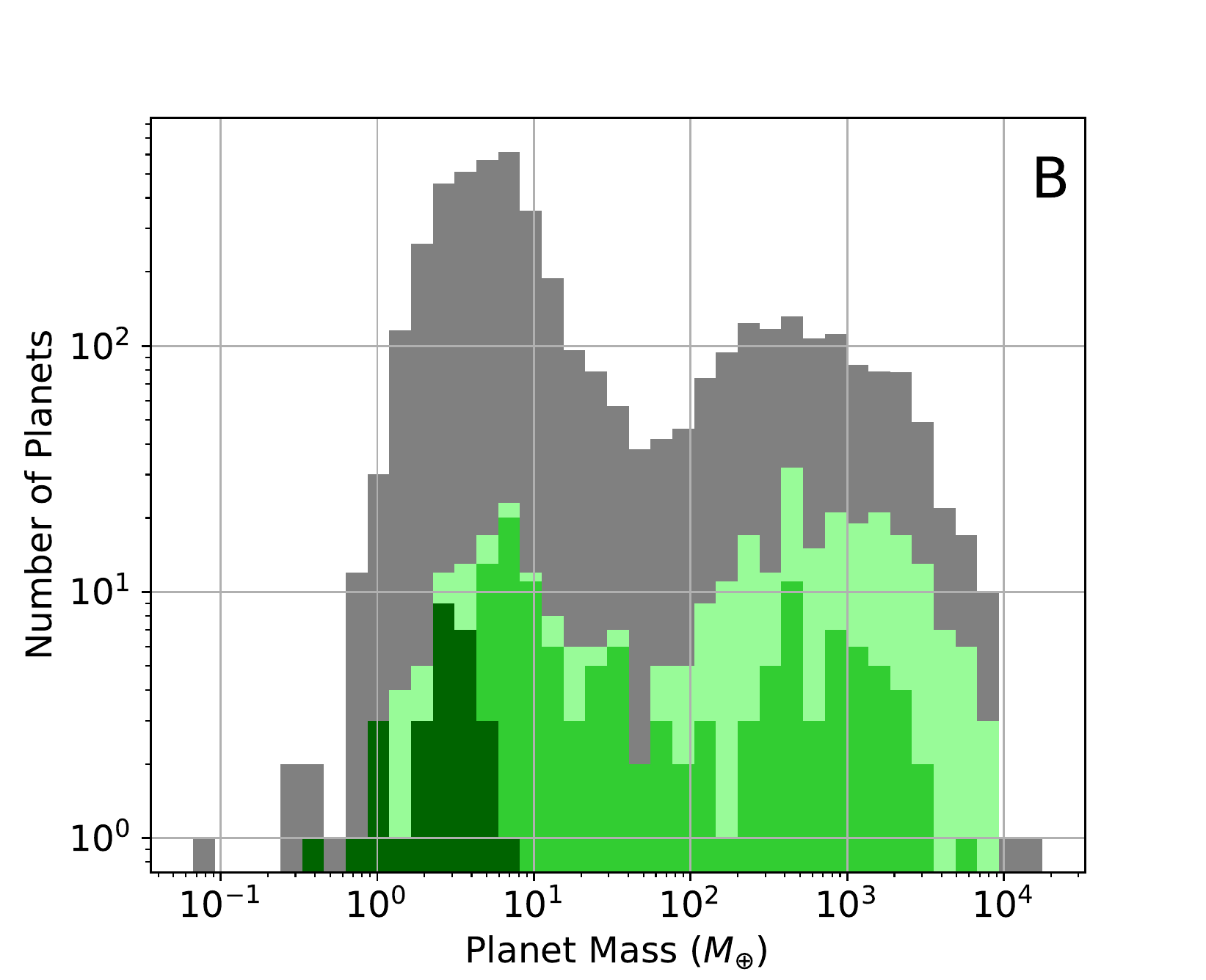}%
}\qquad
\subfloat{%
  \includegraphics[width=1\columnwidth, height=0.8\columnwidth]{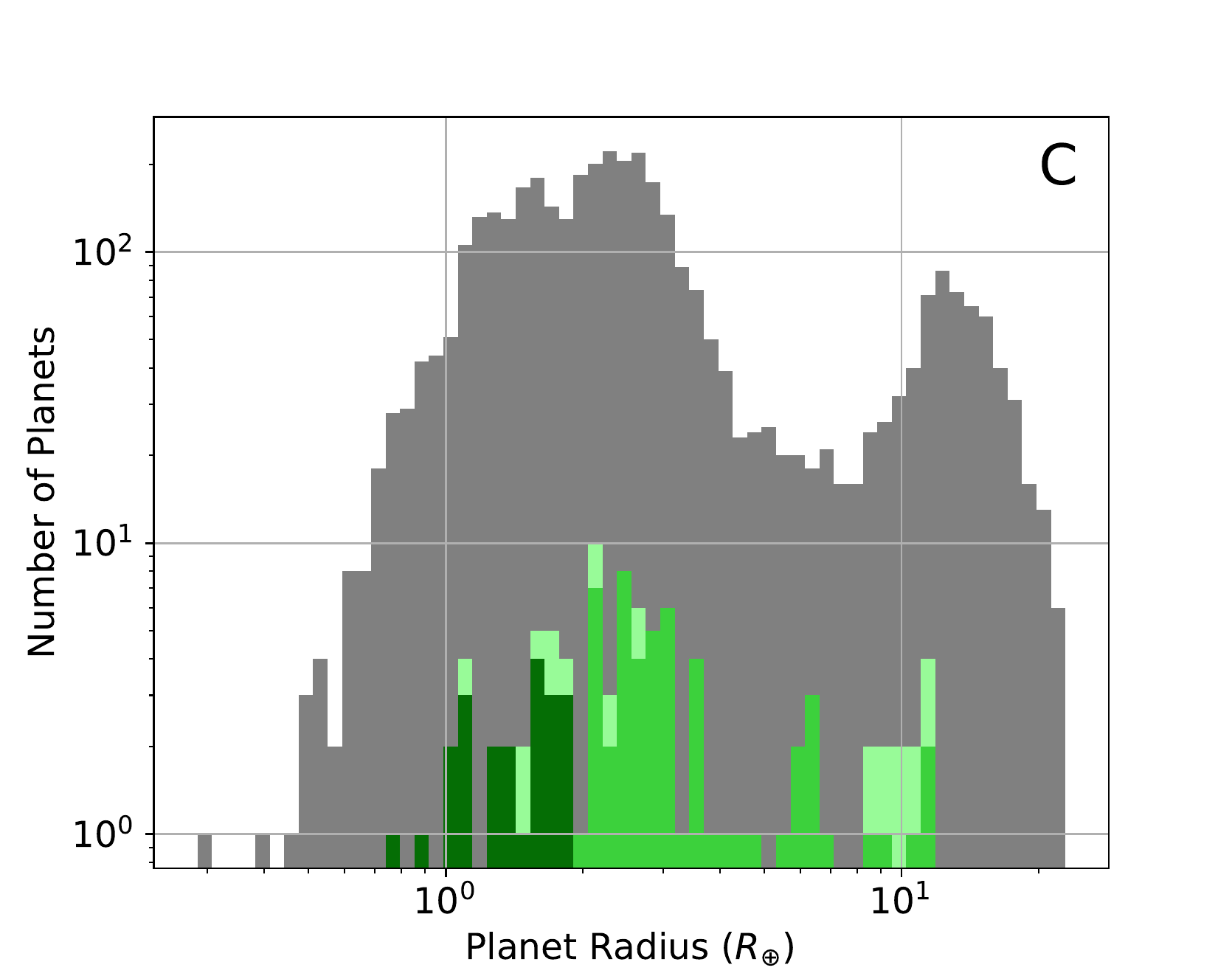}%
}
\subfloat{%
  \includegraphics[width=1\columnwidth, height=0.8\columnwidth]{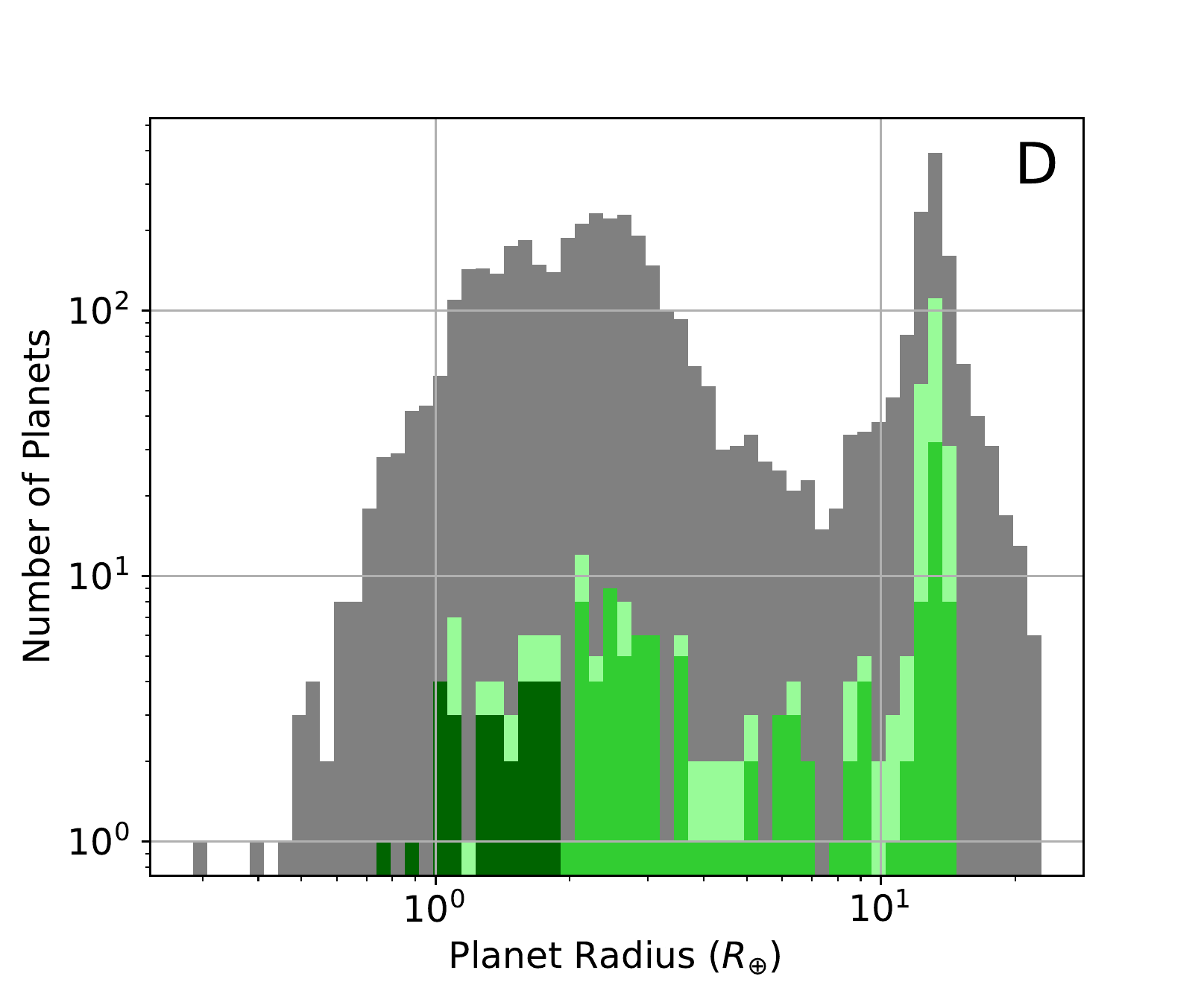}%
}
\caption{Mass and radius histograms of the full catalog of exoplanets from the NEA (gray), planets that spend any time at all in the HZ ($>0$\%HZ) (light green), planets found to orbit in the HZ 100\% of the time (100\%HZ) (medium green), and those $<2R_{\oplus}$ that orbit in the HZ 100\% of the time (Rocky-OHZ) (dark green). Histogram A includes only those planets who have measured values for mass. Histogram B includes both measured values and calculated values of mass using the mass-radius relationship from \citet{chen2017}. Histogram C includes only those planets who have measured values for radius. Histogram D includes both measured values and calculated values of radius.} 
\label{fig:histMR2}
\end{figure*}

\begin{figure*}
\centering 
\vspace{-20pt}
\subfloat{%
  \includegraphics[width=1\columnwidth, height=0.8\columnwidth]{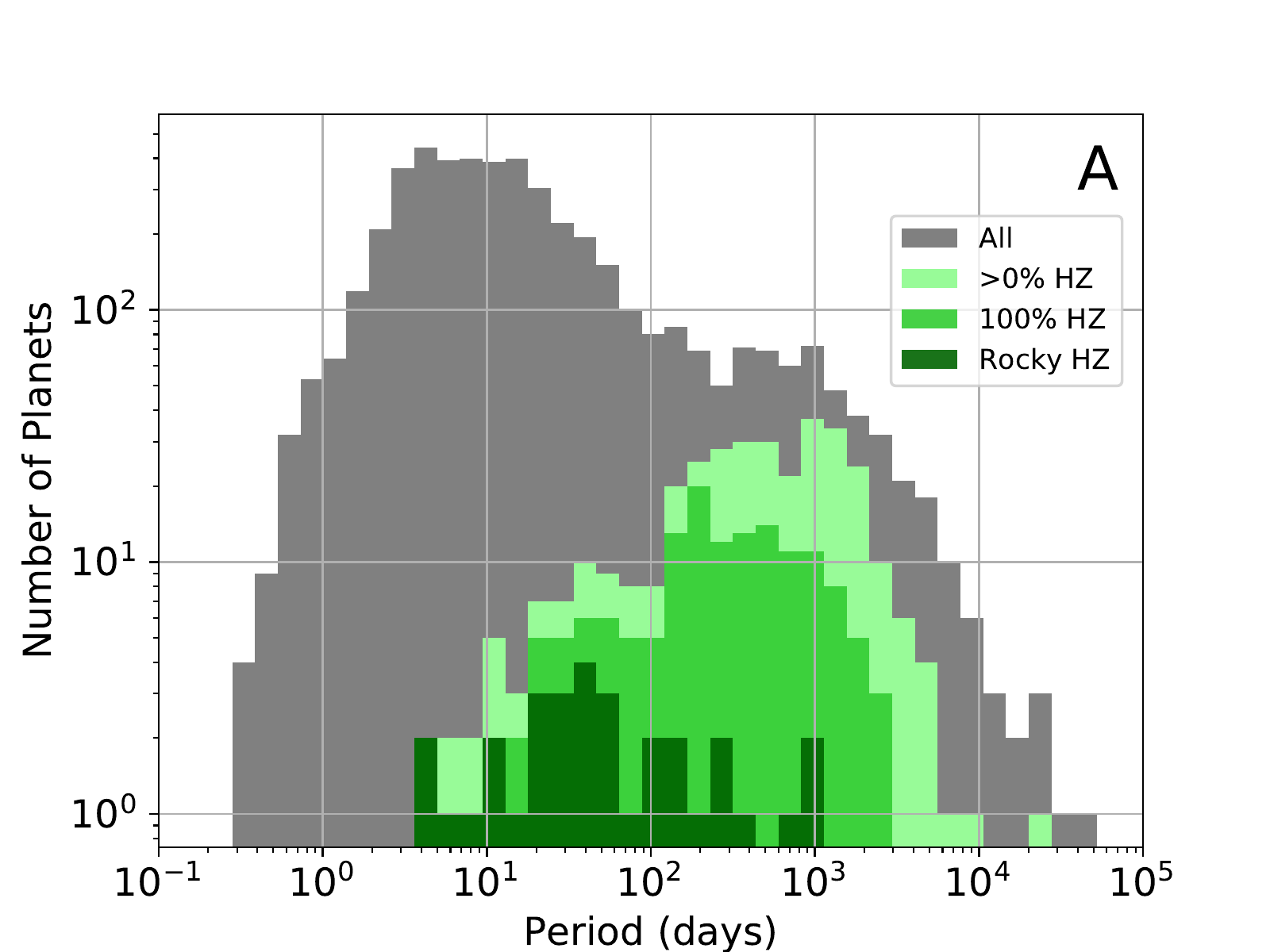}%
}
\subfloat{%
  \includegraphics[width=1\columnwidth, height=0.8\columnwidth]{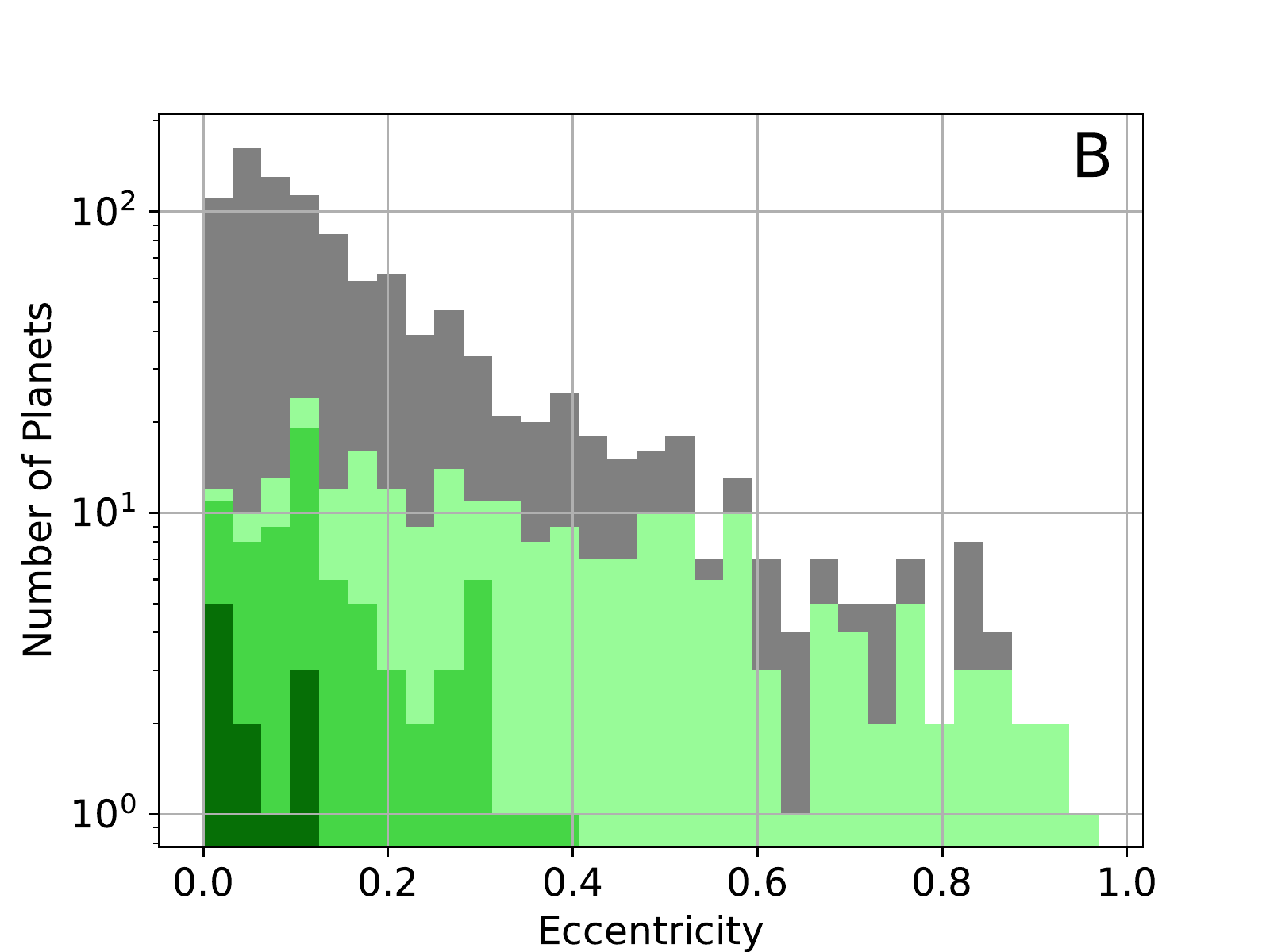}%
}\qquad
\subfloat{%
  \includegraphics[width=1\columnwidth, height=0.8\columnwidth]{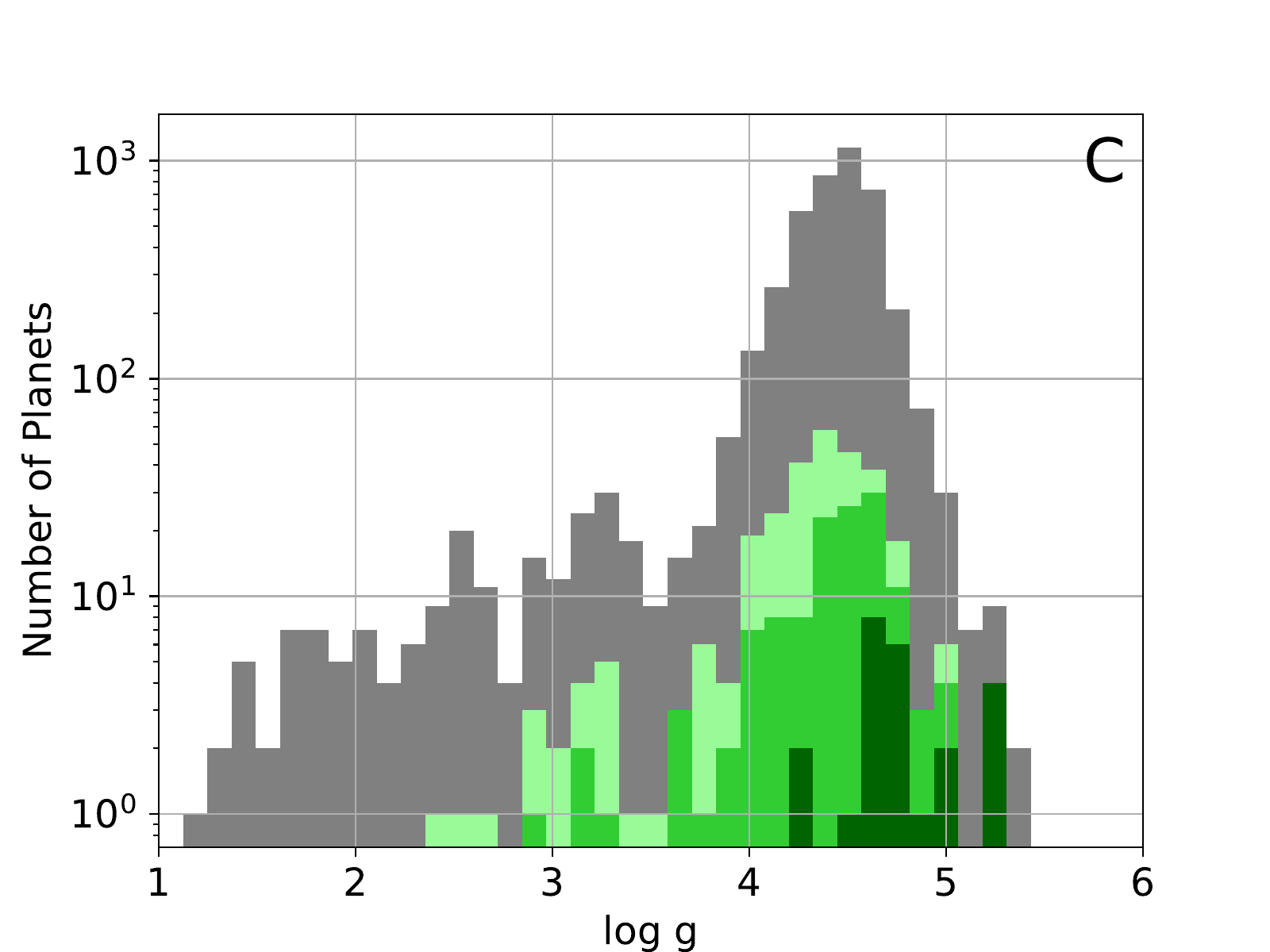}%
}
\subfloat{%
  \includegraphics[width=1\columnwidth, height=0.8\columnwidth]{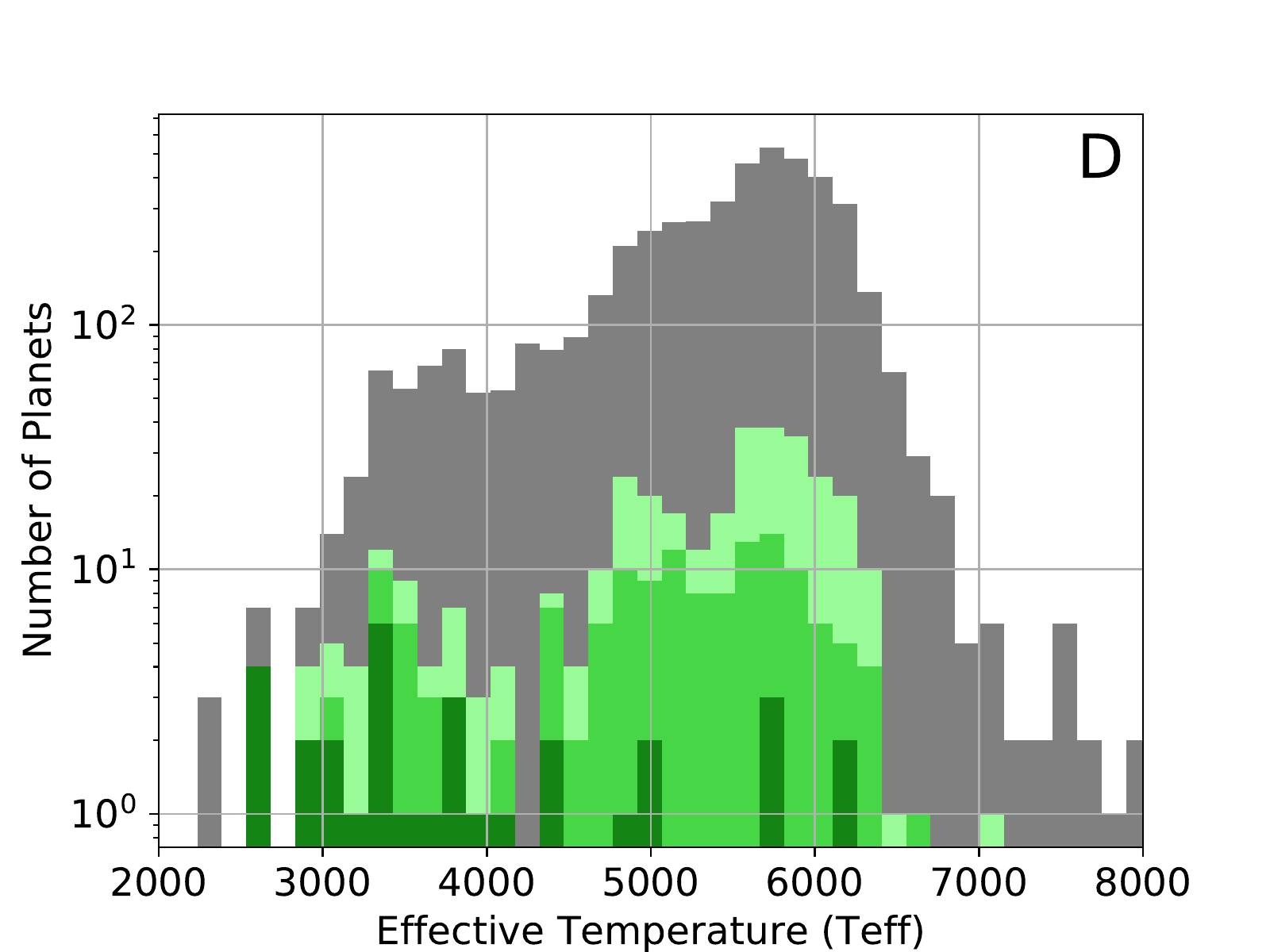}%
}\qquad
\subfloat{%
  \includegraphics[width=1\columnwidth, height=0.8\columnwidth]{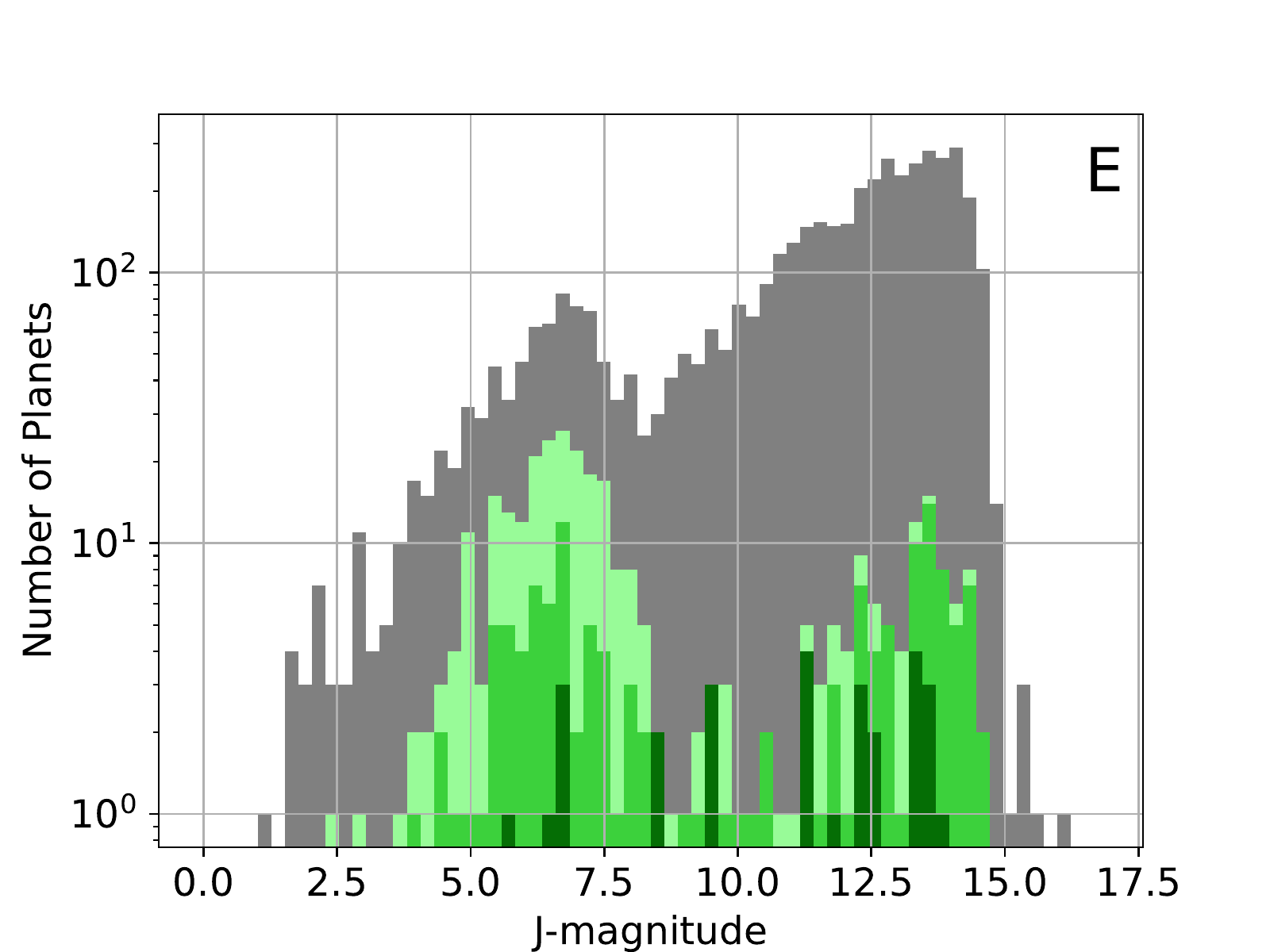}%
}
\subfloat{%
  \includegraphics[width=1\columnwidth, height=0.8\columnwidth]{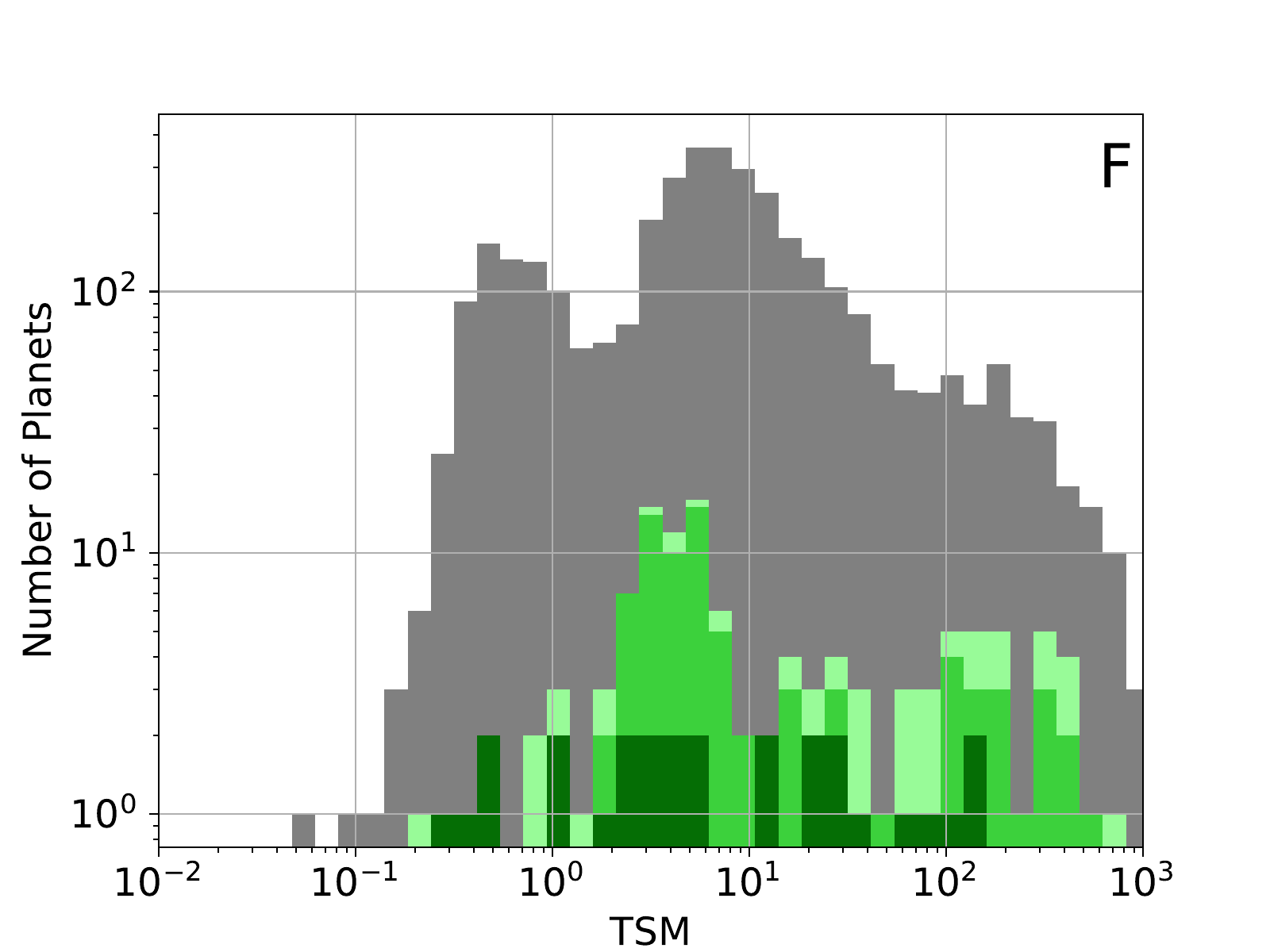}%
}
\caption{ Histogram coloring is the same as Figure \ref{fig:histMR2}. Each histogram includes all planets from the NEA (gray), as well as all planets included in each of the 3 HZ groups (greens). A: Distribution of planet period. The HZ planets peak at higher periods than the full catalog. B: Eccentricity distribution. Only those planets with eccentricity uncertainty measurements are included in these plots. The $<2R_{\oplus}$ planets tend to have more circular orbits. C: $\log~g$ distributions. The HZ planet groups skew right progressively. D: Effective temperature ($T_\mathrm{eff}$) distributions. The $<2R_{\oplus}$ planets tend to be found around stars with lower $T_\mathrm{eff}$. E: $J$ distributions. The $>0$\%HZ that pass through the HZ are more often found around brighter stars. F: TSM distributions. HZ planets have lower TSM values than the full catalog. Note that both transiting and non-transiting planets are included in panel F to represent the overall exoplanet population.} 
\label{fig:hist2}
\end{figure*}

 Figures~\ref{fig:histMR2} and \ref{fig:hist2} show distributions of planet and stellar parameters for the entire exoplanet catalog (gray), the $>0\%$HZ planets (light green), planets that spend 100\% of the time in the OHZ ($100\%$HZ; medium green), and planets $<2$~$R_{\oplus}$ that spend 100\% of the time in the OHZ (Rocky-OHZ; dark green). As of printing 4586 planets had sufficient data to be included in the full catalog, 328 planets in the $>0\%$HZ subset, 143 planets in the $100\%$HZ subset, and 29 planets in the Rocky-OHZ subset.
 In order to directly compare the groups, the bin sizes are the same for each group. Sturges' rule was applied to the full catalog group to determine the appropriate histogram bin sizes and then used for all groups.
 
\begin{figure*}
\centering
\subfloat{%
  \includegraphics[width=1\columnwidth, height=0.8\columnwidth]{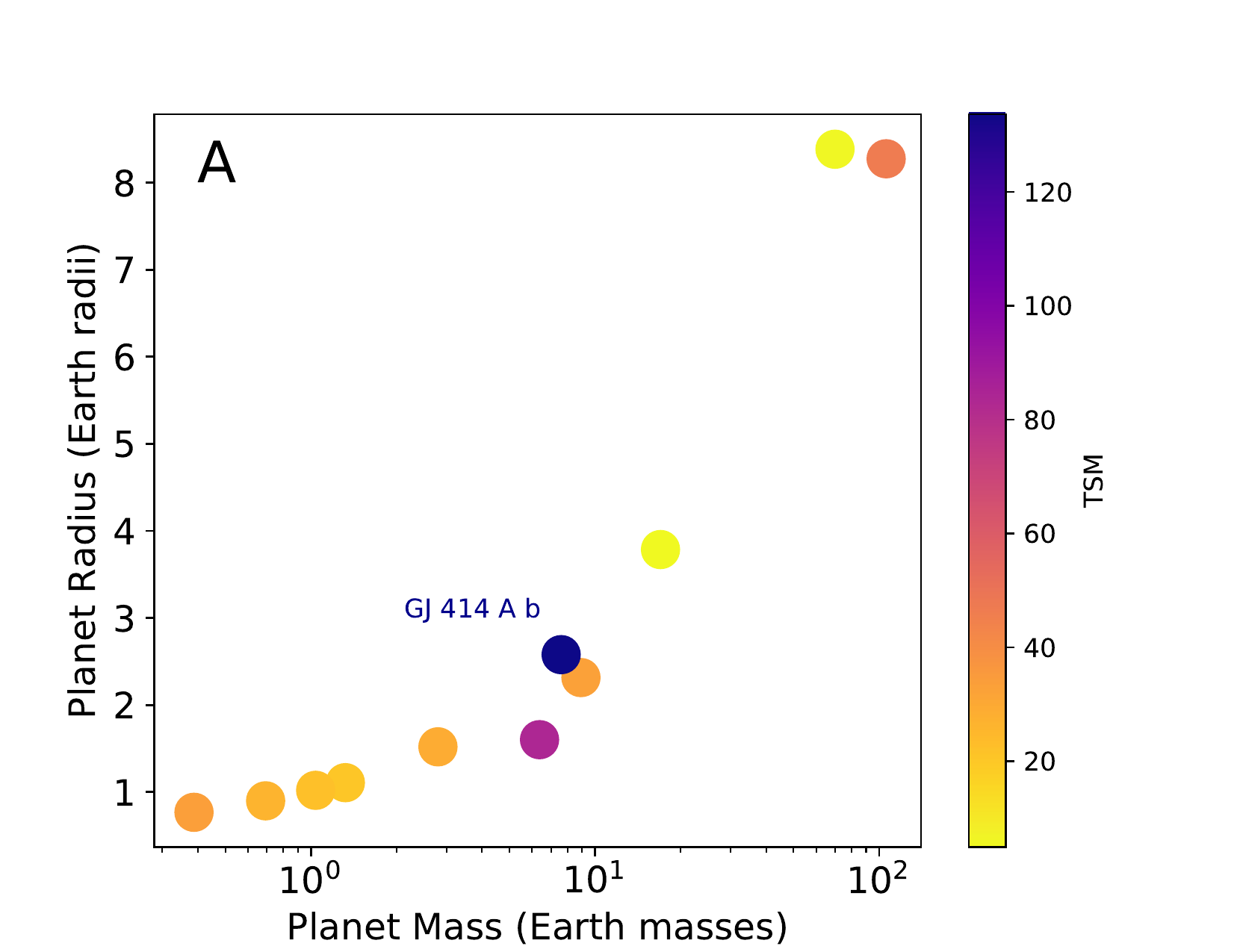}%
}
\subfloat{%
  \includegraphics[width=1\columnwidth, height=0.8\columnwidth]{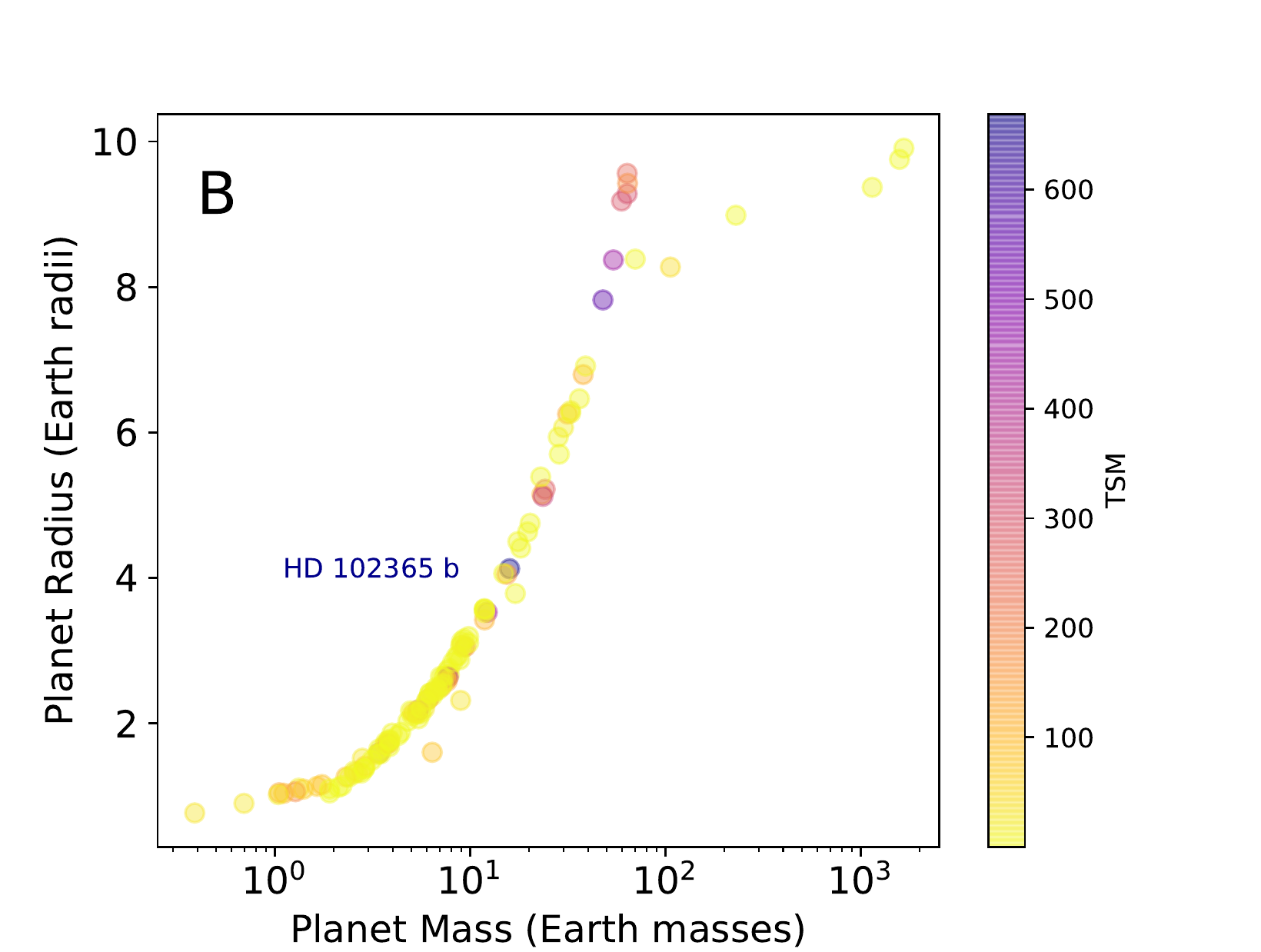}%
}\qquad
\subfloat{%
  \includegraphics[width=1\columnwidth, height=0.8\columnwidth]{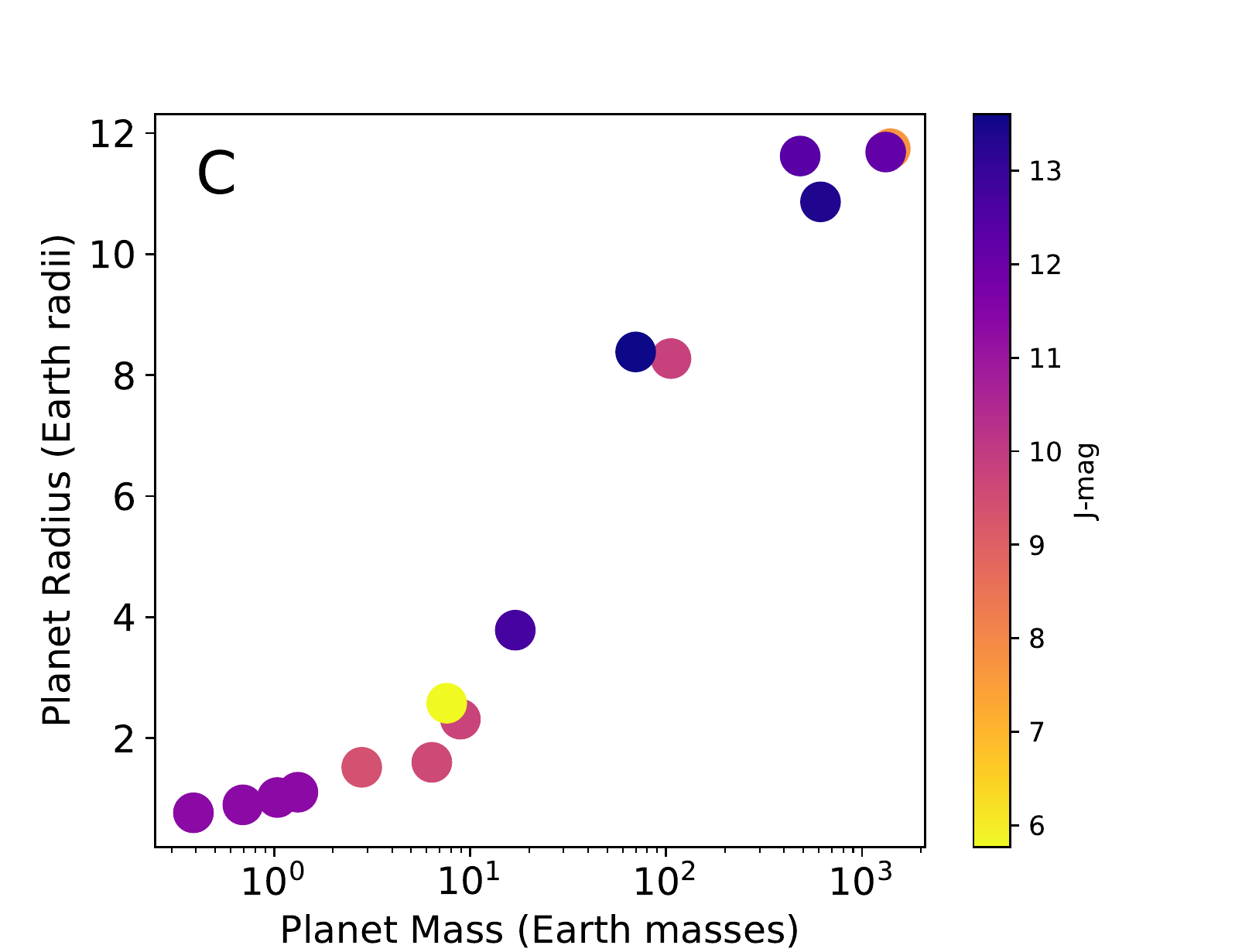}%
}
\subfloat{%
  \includegraphics[width=1\columnwidth, height=0.8\columnwidth]{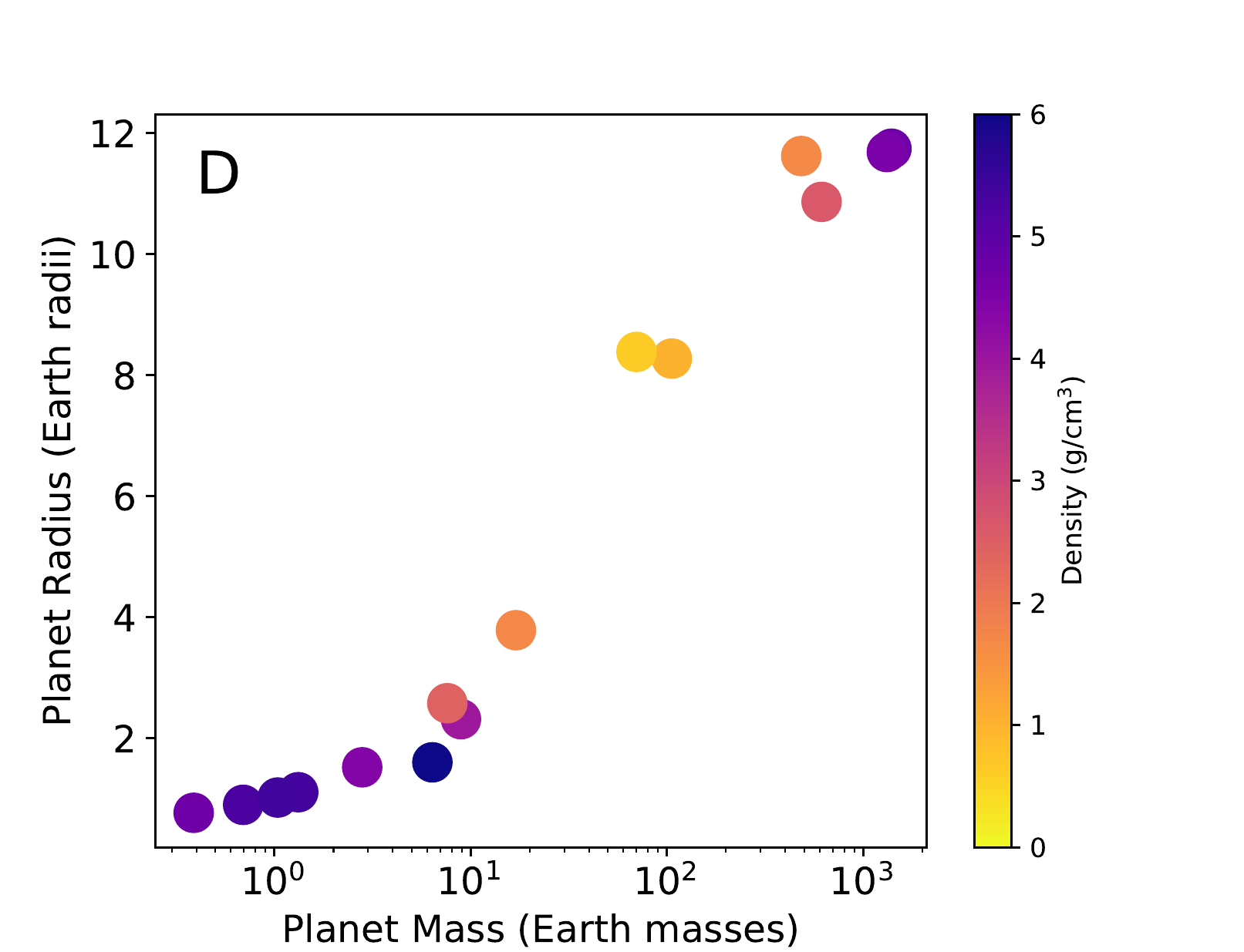}%
}
\caption{Panels A, C and D include only the planets found in the HZ that have both a measured mass and radius. Panel B also includes mass or radii calculations from \citet{chen2017}. A and B: Color of data points indicate transmission spectroscopy metric (TSM) of the planet \citep{kempton2018}. The planets with the highest TSM value in each plot are labelled. GJ~414~A~b has the greatest TSM of all known transiting planets, including those without mass measurements. HD~102365~b has the greatest TSM of all HZ planets though, like many of the planets included in Figure 4B, it is not known to transit. C: Color of data points indicate $J$ as indicated by the color bar on the right. The brightness of GJ~414~A contributed to the high TSM value. D: Color of data points indicate bulk density of the planet in $g/cm^3$. The planets with the greatest density on the bottom left are the TRAPPIST-1 planets. 
\label{fig:MRscatter}}
\end{figure*}

Scatter plots including measured values of mass and radius alongside those with \citet{chen2017} values can be found in Figure \ref{fig:MRscatter}. Each of these is color coded with a third parameter: TSM, $J$ or density. 
 
\begin{figure*}
\centering 
\subfloat{%
  \includegraphics[width=1\columnwidth, height=0.8\columnwidth]{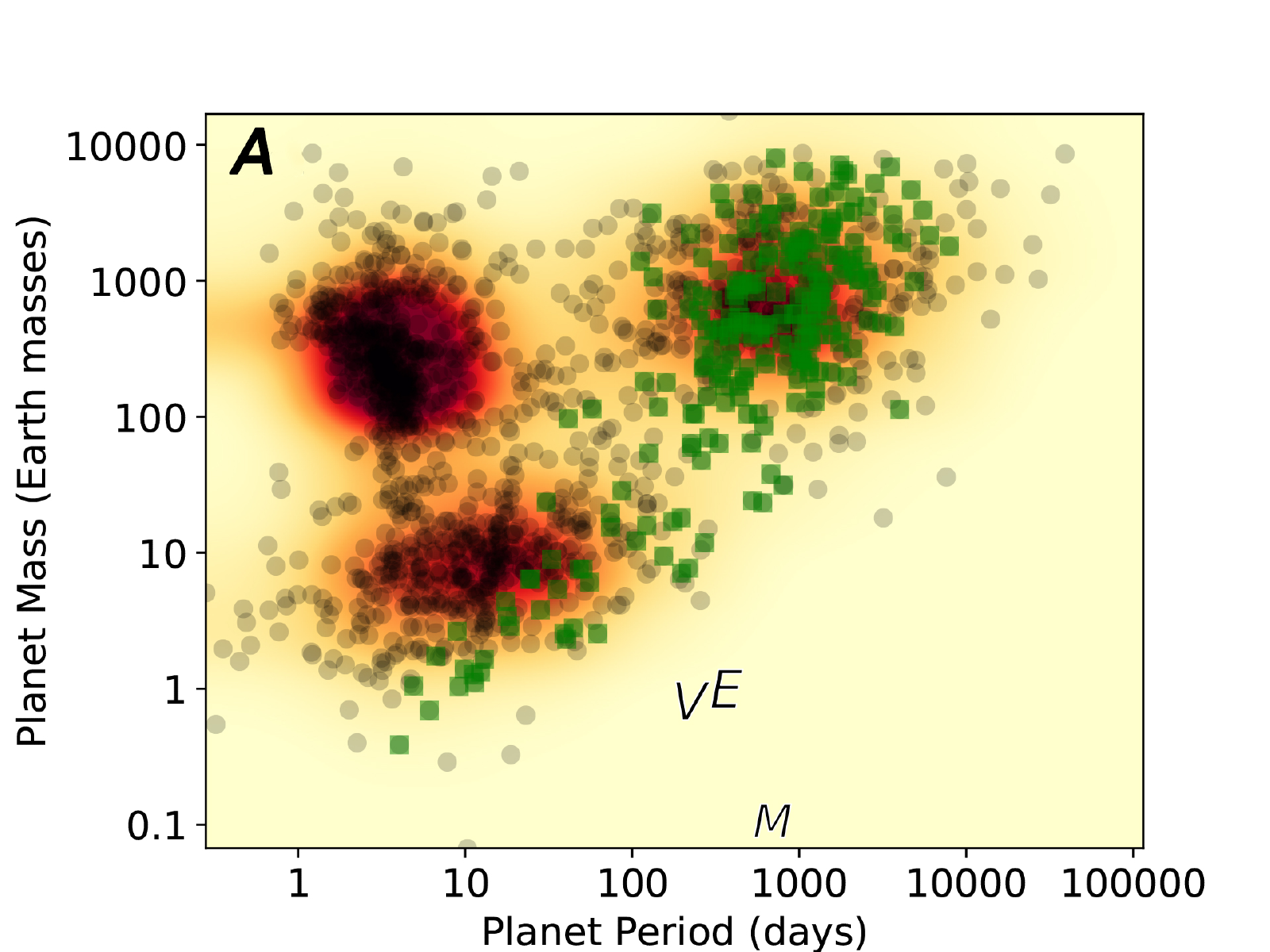}%
}
\subfloat{%
  \includegraphics[width=1\columnwidth, height=0.8\columnwidth]{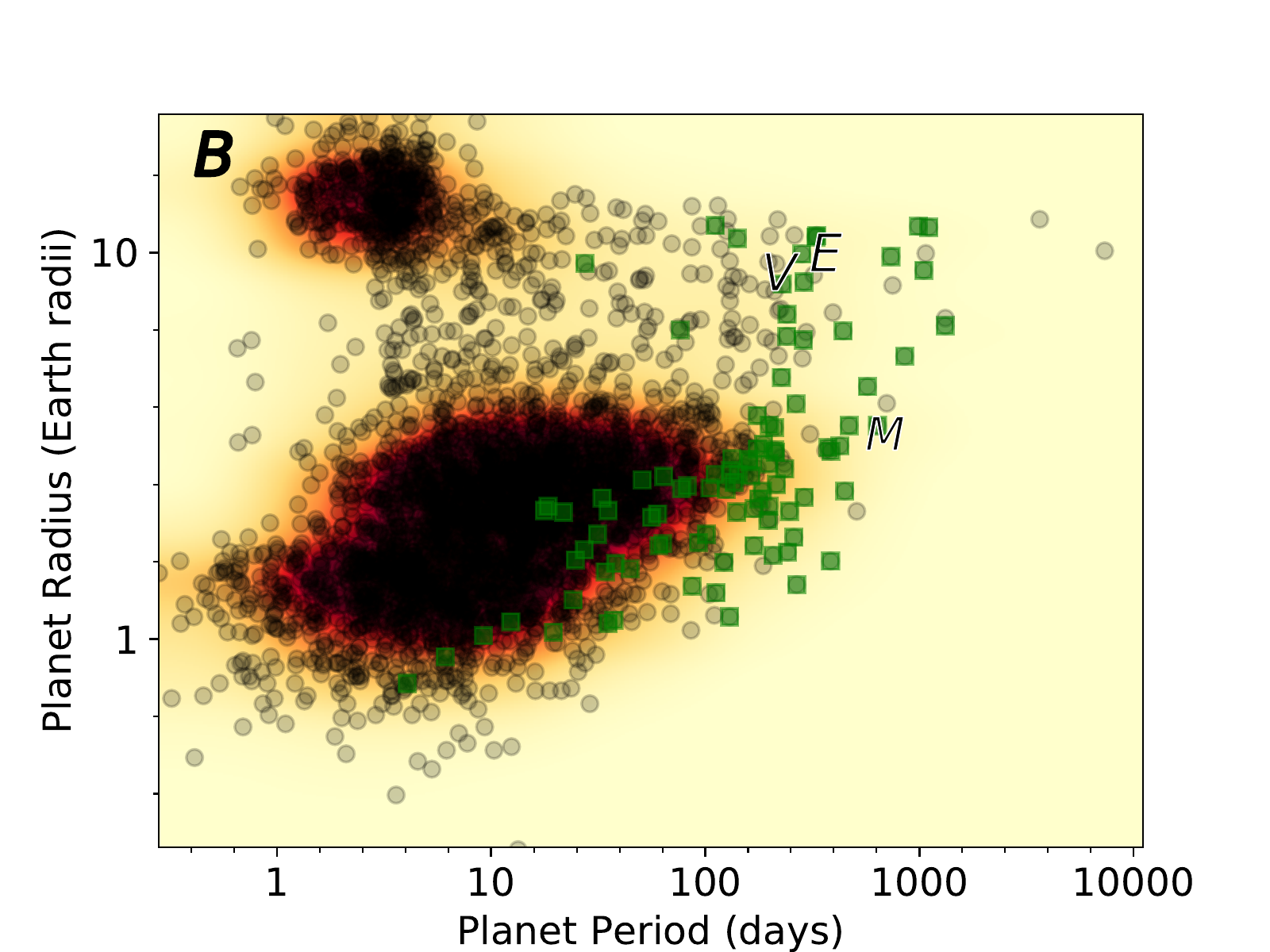}%
}\qquad
\subfloat{%
  \includegraphics[width=1\columnwidth, height=0.8\columnwidth]{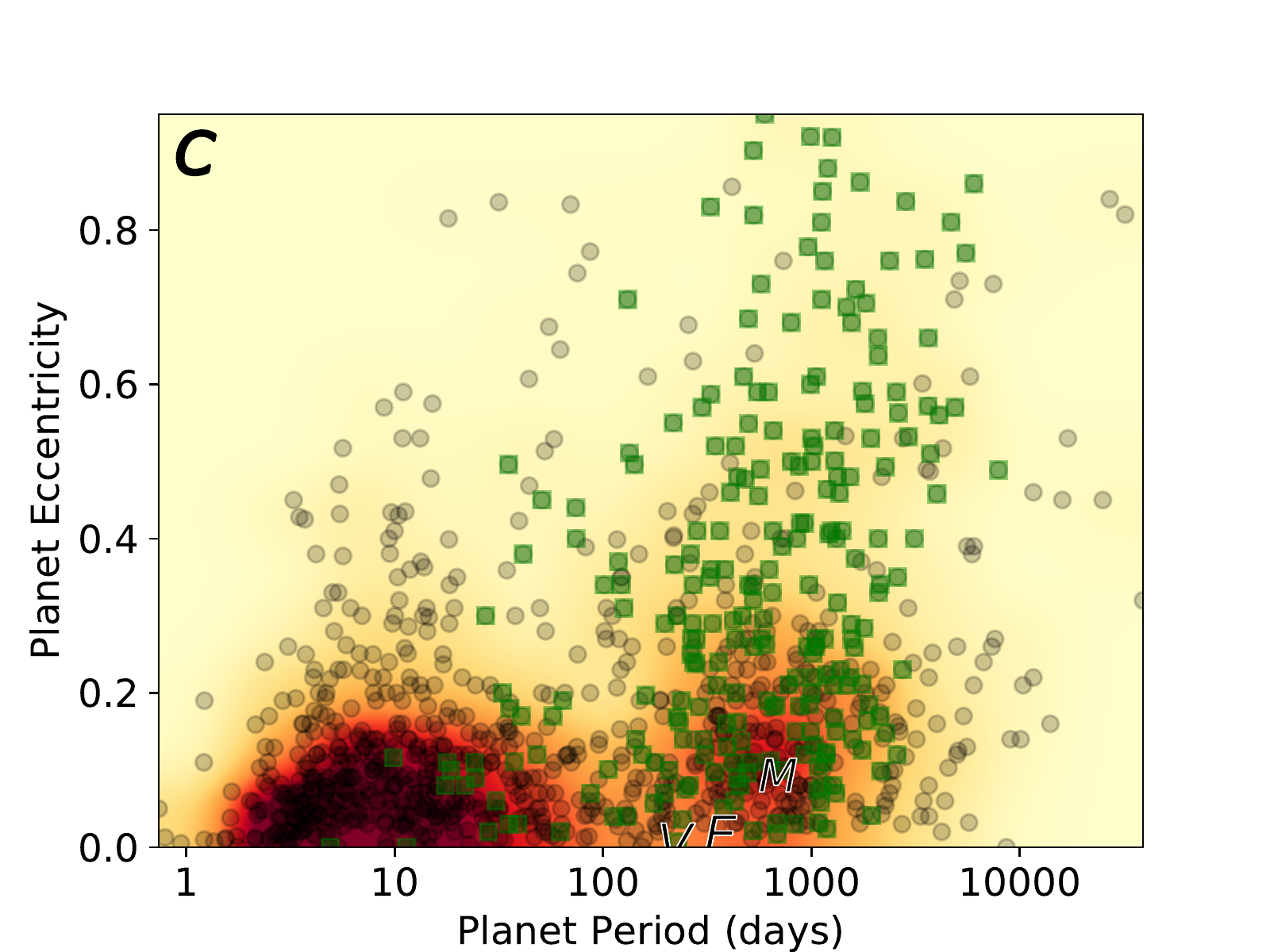}%
}
\subfloat{%
  \includegraphics[width=1\columnwidth, height=0.8\columnwidth]{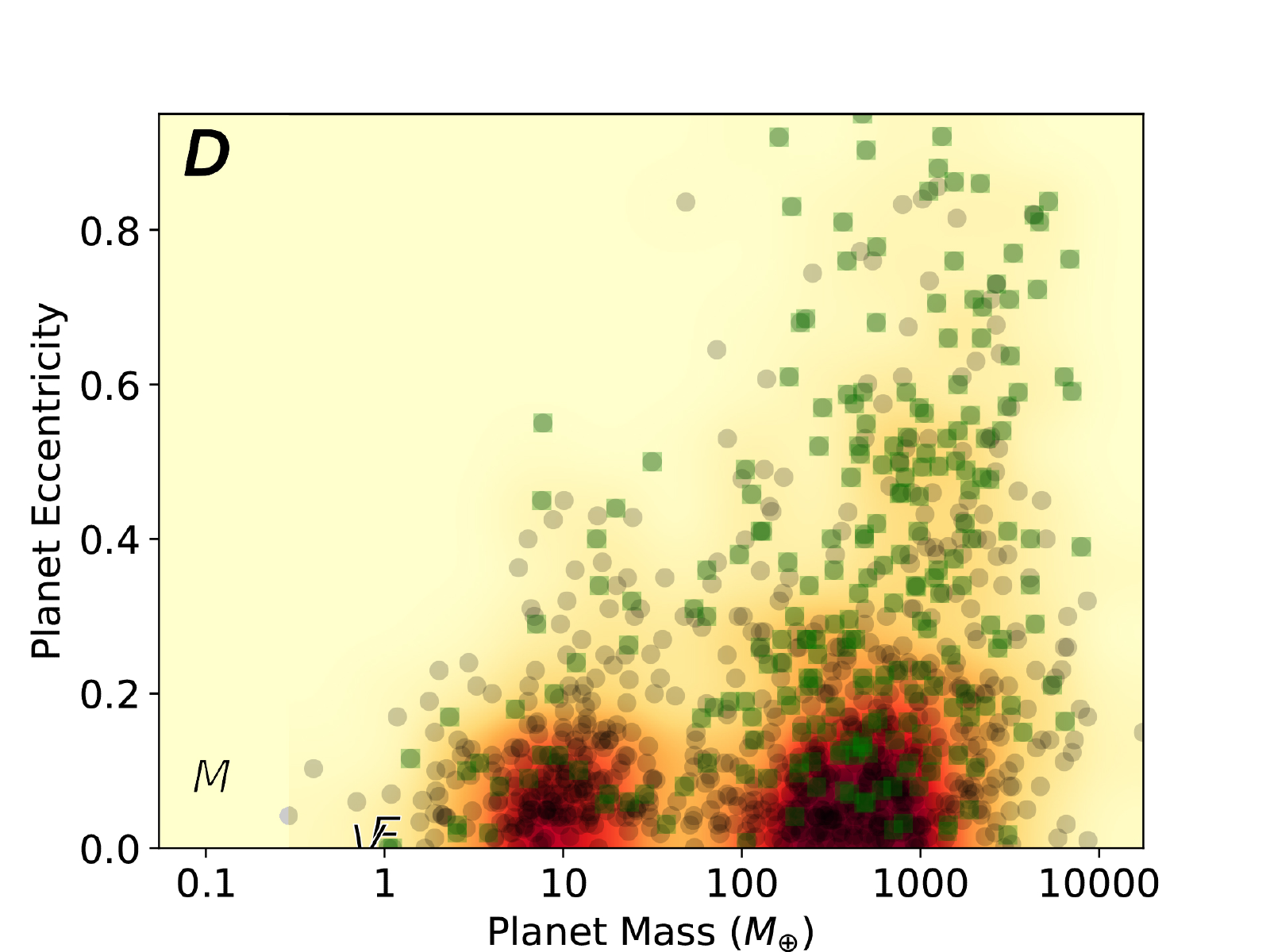}%
}\qquad
\subfloat{%
  \includegraphics[width=1\columnwidth, height=0.8\columnwidth]{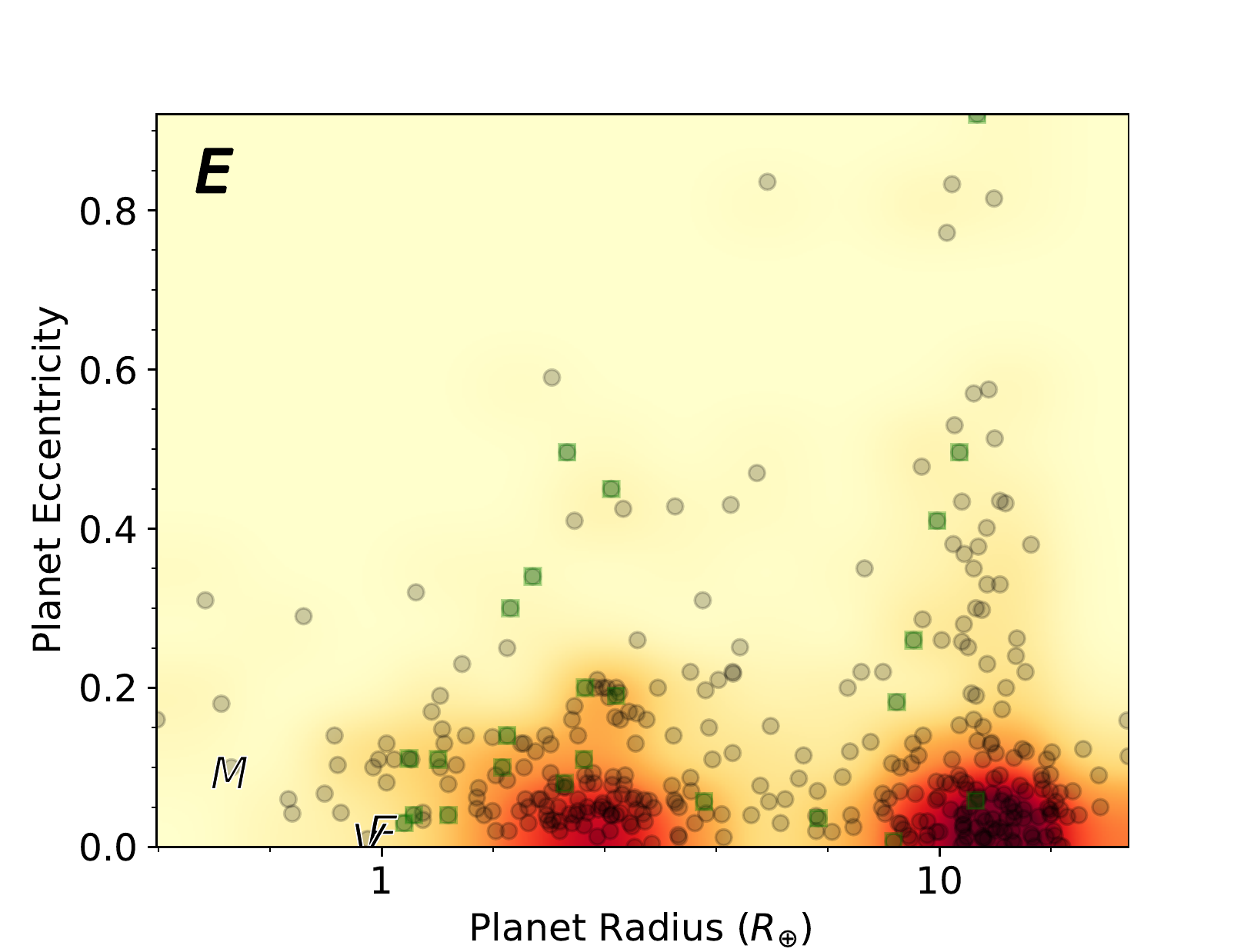}%
}
\subfloat{%
  \includegraphics[width=1\columnwidth, height=0.8\columnwidth]{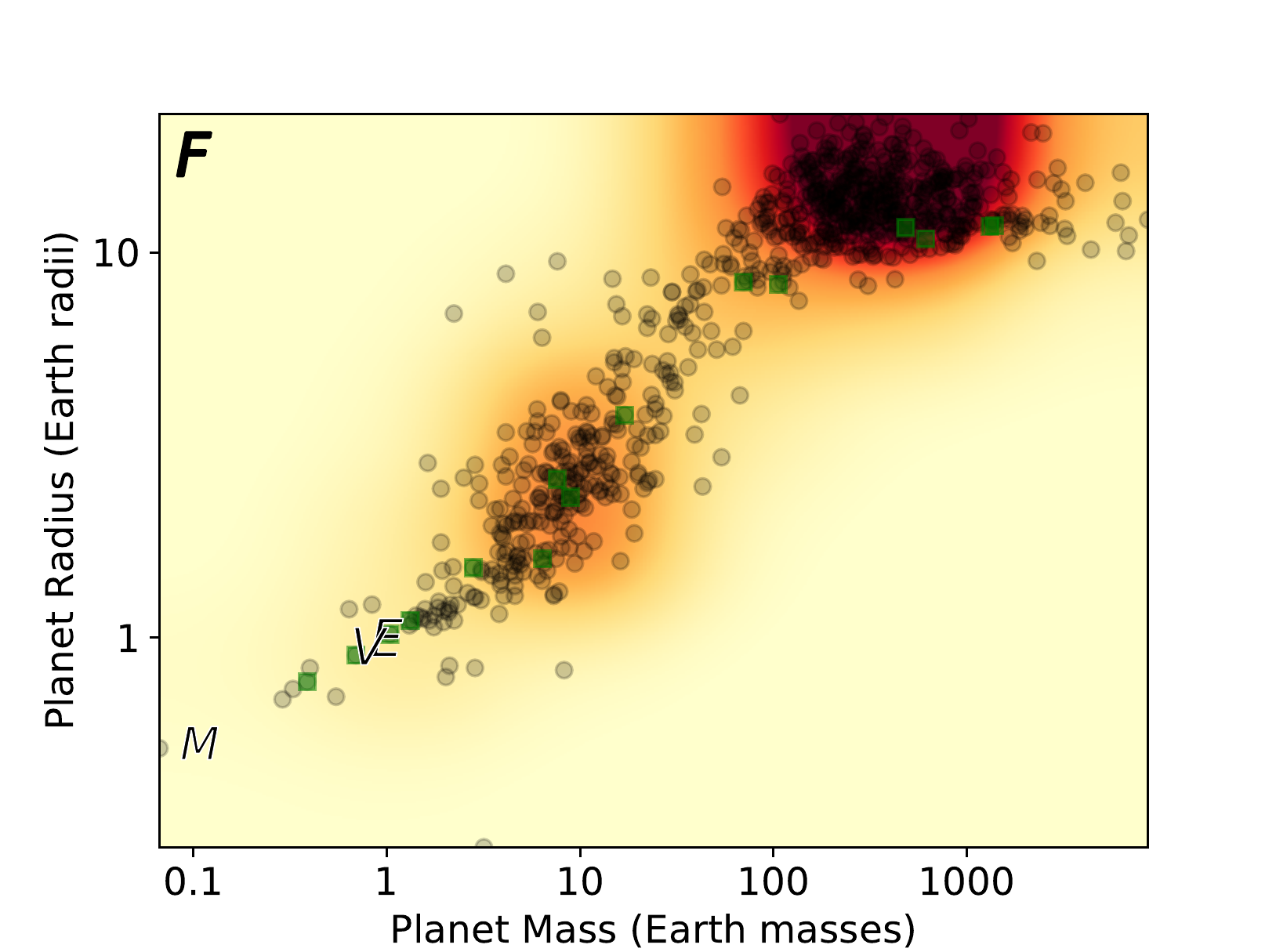}%
}
\caption{Heat maps with HZ planets (green) overlaying the full catalog of known exoplanets (gray). Top left: Compared to the full catalog, HZ planets tend to have longer periods and larger masses. Top right: The planet period vs radius heat map highlights the gap between the mini-Neptune sized planets and the larger Jupiter sized planets. Middle left: Longer period planets tend to have a wider distribution of eccentricity in both the full catalog and HZ planets. Middle right: So too do the larger mass planets. Bottom left: This trend of larger planets having a broader range of eccentricities is not seen in the HZ planets in the Radius-eccentricity plot, but this may be due to a smaller population. Bottom right: The measured mass versus radius plot reveals clusters of Jupiter-sized planets and Super Earths in the full catalog but not in the HZ planets. 
\label{fig:heat1}}
\end{figure*}

\begin{figure}
  \begin{center}
    \includegraphics[width=0.45\textwidth]{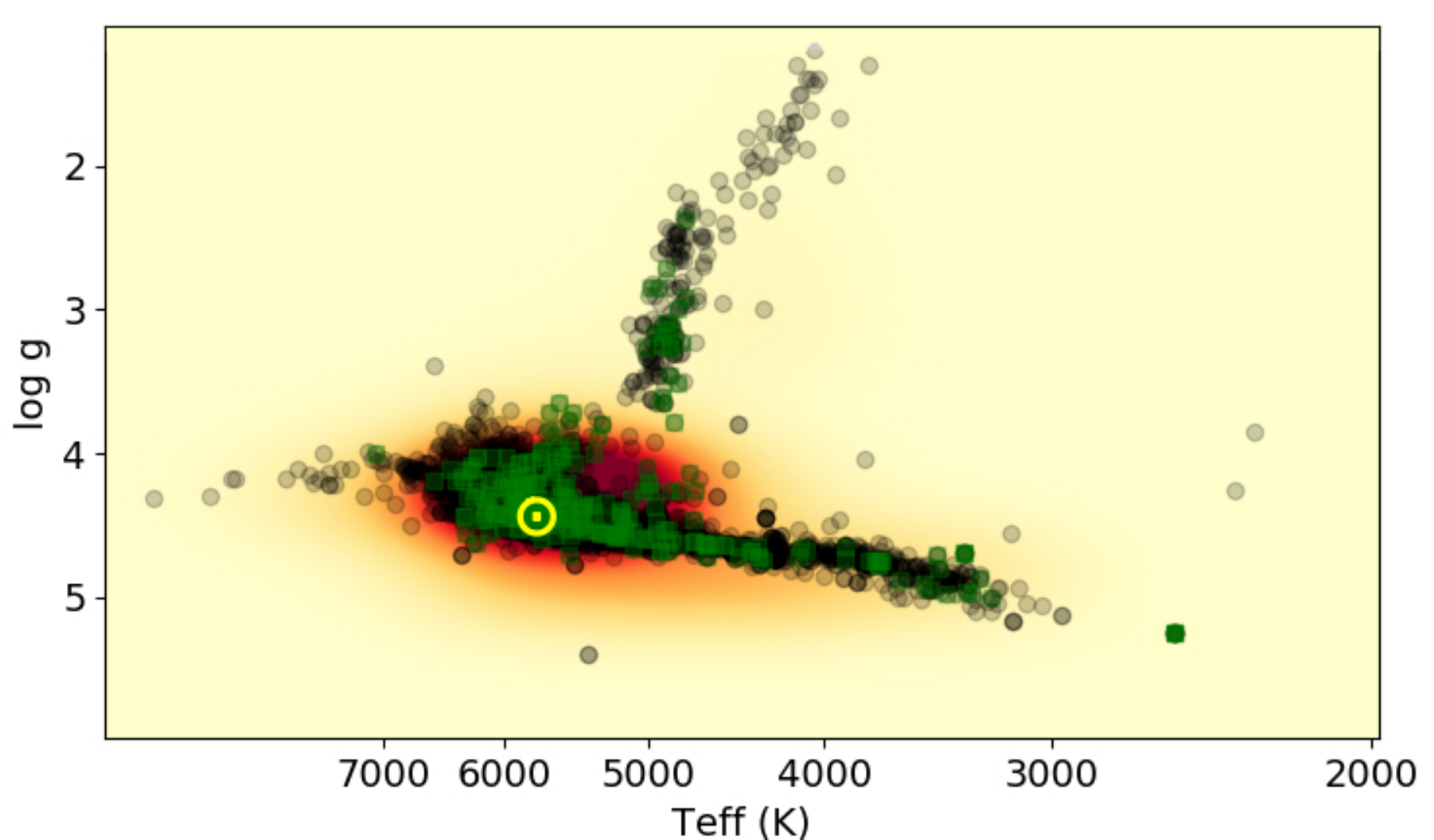}
  \end{center}
  \caption[$\log~g$ vs $T_\mathrm{eff}$]{Heat maps of stellar effective temperature and $\log~g$. The population of $>0\%$HZ planets span the same distribution of $T_\mathrm{eff}$ and $\log~g$ as the full catalog of planets, except for amongst the more evolved stars at the top of the plot. These stars either may not host planets in the HZ, or the properties of these stars may mean detection of any HZ planets is beyond current capabilities. 
 }
  \label{fig:heat2}
\end{figure}

Figures \ref{fig:heat1} and \ref{fig:heat2} include scatter plots of the entire exoplanet catalog (gray) versus the $>0\%$HZ planets (green). These plots also include a heat map to allow easy identification of clusters and their relative density. 

\floattable
\startlongtable
\begin{deluxetable*}{lllllllllllllll}
  \tablewidth{0pc}
  \setlength{\tabcolsep}{2pt}
 \tabletypesize{\scriptsize}
  \centering
  \tablecaption{Habitable Zone Planet Properties.
  \label{tab:HZpl}}
\tablehead{
\colhead{Name}  & \colhead{$M_p$ \tablenotemark{+}} & \colhead{$R_p$ \tablenotemark{+}} &  \colhead{$P$}      &  \colhead{$a$  }   &  \colhead{$e$ } &  \colhead{Flux} &  \colhead{R1 } &  \colhead{R2 }    &  \colhead{R3 } &  \colhead{R4 } &  \colhead{ CHZ} &  \colhead{ OHZ} &  \colhead{TSM\tablenotemark{*}} & \colhead{$K$}\\
\colhead{} & \colhead{$M_\oplus$} & \colhead{$R_\oplus$} & \colhead{days}  & \colhead{AU} & \colhead{} & \colhead{$W/m^2$} & \colhead{AU} & \colhead{AU} & \colhead{AU} & \colhead{AU} & \colhead{$\%$} & \colhead{$\%$}  & \colhead{} & \colhead{$m/s$}}
\startdata
GJ 3293 d & 7.6$^{+1.05}_{-1.05}$ & \textit{2.67} & 48.13 & 0.194 & 0.12 & 797.6 & 0.121 & 0.154 & 0.299 & 0.315 & 100 & 100 & 145.1 & 2.4 \\
GJ 357 d & 6.1$^{+1}_{-1}$ & \textit{2.33} & 55.66 & 0.204 & 0 & 522.3 & 0.103 & 0.131 & 0.254 & 0.268 & 100 & 100 & 168.6 & 2.1 \\
GJ 667 C c & 3.81$^{+1.5}_{-1.2}$ & \textit{1.73} & 28.14 & 0.125 & 0.02 & 1197.6 & 0.096 & 0.122 & 0.238 & 0.251 & 100 & 100 & 150.7 & 1.7 \\
GJ 667 C e (!) & 2.54$^{+1.6}_{-1.4}$ & \textit{1.31} & 62.24 & 0.213 & 0.02 & 412.5 & 0.096 & 0.122 & 0.238 & 0.251 & 100 & 100 & 11.4 & 0.9 \\
GJ 667 C f (!) & 2.54$^{+1.4}_{-1.2}$ & \textit{1.33} & 39.03 & 0.156 & 0.03 & 768.9 & 0.096 & 0.122 & 0.238 & 0.251 & 100 & 100 & 13.9 & 1.0 \\
GJ 682 b & 4.4$^{+3.7}_{-2.4}$ & \textit{1.89} & 17.48 & 0.080 & 0.08 & 426.5 & 0.037 & 0.047 & 0.093 & 0.098 & 100 & 100 & -- & 2.6 \\
HD 10442 b & 472.61$^{+24.16}_{-24.16}$ & \textit{13.46} & 1032.30 & 2.010 & 0.132 & 685.4 & 1.120 & 1.419 & 2.577 & 2.718 & 100 & 100 & -- & 29.9 \\
HD 106270 b & 3219.62$^{+85.81}_{-85.81}$ & \textit{12.3} & 1888.00 & 3.340 & 0.185 & 716.0 & 1.844 & 2.336 & 4.152 & 4.379 & 100 & 100 & -- & 135.5 \\
\enddata
\tablenotetext{+}{Values in italics indicate calculated values from \citet{chen2017}}
\tablenotetext{(!)}{ Planet is flagged as controversial in the NEA.}
\tablenotetext{*}{TSM values where planet radii is calculated from \citet{chen2017} are provided for future use in the case where the planet is later found to transit.}
\end{deluxetable*}

\floattable
\startlongtable
\begin{deluxetable*}{llllllll}
  \tablewidth{0pc}
 \tabletypesize{\scriptsize}
  \centering
  \tablecaption{Stellar Properties.
  \label{tab:HZstar}}
\tablehead{
\colhead{Star Name}  & \colhead{Stellar Mass} & \colhead{Stellar Radius} &  \colhead{Eff Temp}      &  \colhead{$\log g$}     &  \colhead{Luminosity} &  \colhead{$J$}  &  \colhead{$V$} \\
\colhead{} & \colhead{$M_\odot$} & \colhead{$R_\odot$} & \colhead{K} & \colhead{dex} & \colhead{$L_\odot$} & \colhead{}  & \colhead{}}
\startdata
16 Cyg B & 1.08 & 1.13 & 5750 & 4.36 & 1.25 & 4.993 & 6.215 \\
30 Ari B & 1.93 & 1.41 & 6331 & 4.43 & 2.87 & 6.080 & 7.092 \\
47 UMa & 1.06 & 1.21 & 5872 & 4.3 & 1.58 & 3.960 & 5.034 \\
55 Cnc & 0.91 & 0.94 & 5172 & 4.43 & 0.57 & 4.768 & 5.951 \\
BD-00 4475 & 0.81 & 1.12 & 5040 & 4.25 & 0.72 & 7.483 & 9.010 \\
BD-06 1339 & 0.7 & 0.59 & 4324 & 4.74 & 0.10 & 7.128 & 9.700 \\
BD-08 2823 & 0.74 & 1.23 & 4746 & 4.13 & 0.68 & 7.960 & 9.860 \\
BD-11 4672 & 0.65 & 0.64 & 4550 & 4.64 & 0.16 & 7.651 & 10.030 \\
\enddata
\tablenotetext{(!)}{ Planet is flagged as controversial in the NEA.}
\end{deluxetable*}

\floattable
\startlongtable
\begin{deluxetable*}{lllllllllllllllll}
  \tablewidth{0pc}
 \tabletypesize{\scriptsize}
  \centering
  \tablecaption{Circumbinary HZ Planet Properties.
  \label{tab:cb}}
\tablehead{
\colhead{Planet Name}  & \colhead{$M_p$ \tablenotemark{+}} & \colhead{$R_p$} &  \colhead{$P$}      &  \colhead{$a$  }   &  \colhead{$e$ } & 
\colhead{$M_{\star,\mathrm{A}}$} &  \colhead{$R_{\star,\mathrm{A}}$}    &  \colhead{$T_\mathrm{eff},\mathrm{A}$ } &  \colhead{$\log g_\mathrm{A}$ } &  \colhead{$L_{\star,\mathrm{A}}$} &  \colhead{$J_\mathrm{A}$} &  \colhead{$M_{\star,\mathrm{B}}$}  &  \colhead{$R_{\star,\mathrm{B}}$} \\
\colhead{} & \colhead{$M_\oplus$} & \colhead{$R_\oplus$} & \colhead{days}  & \colhead{AU} & \colhead{} 
& \colhead{$M_\odot$} & \colhead{$R_\odot$} & \colhead{K} & \colhead{dex} & \colhead{$L_\odot$} & \colhead{} & \colhead{$M_\odot$} & \colhead{$R_\odot$} }
\startdata
Kepler-16 b & 105.84$^{+5.09}_{-5.09}$ & 8.27$^{+0.029}_{-0.029}$ & 228.78 & 0.705 & 0.0069
& 0.69 & 0.65 & 4450 & 4.65 & 0.15 & 9.815 & 0.2 & 0.23 \\
Kepler-47 c & 3.17$^{+2.18}_{-1.25}$ & 4.65$^{+0.09}_{-0.07}$ & 303.23 & 0.9638 & 0.044 & 0.99 & 	0.93 & 5611 & 4.50 & 0.89 & 13.97 & 0.85 & 0.87 \\
Kepler-453 b & \textit{31.95} & 6.07$^{+0.039}_{-0.039}$ & 240.50 & 0.790 & 0.0359 
& 0.94 & 0.83 & 5527 & 4.57 & 0.58 & 12.349 & 0.86 & 0.77 \\
Kepler-1647 b & 483.00$^{+206}_{-206}$ & 11.62$^{+0.138}_{-0.138}$ & 1107.59 & 2.721 & 0.0581 
& 1.22 & 1.79 & 6210 & 4.02 & 4.28 & 12.376 & 0.882 & 1.464 \\
Kepler-1661 b & 17.00$^{+12}_{-12}$ & 3.79$^{+0.06}_{-0.06}$ & 175.06 & 0.633 & 0.057 
& 0.84 & 0.76 & 5100 & 4.66 & 0.35 & 12.729 & 0.26 & 0.28 \\
\enddata
\tablenotetext{+}{Values in italics indicate calculated values from \citet{chen2017}}
\end{deluxetable*}

\setlength{\belowcaptionskip}{2pt}

\setlength{\belowcaptionskip}{2pt}

\setlength{\belowcaptionskip}{2pt}

\setlength{\belowcaptionskip}{2pt}

\section{Discussion}
\label{discussion}


\subsection{Demographics}
\label{demo}

The mass and radius histograms of the Rocky-OHZ planets (dark green), 100\%HZ planets (medium green), $>0\%$HZ planets (light green), and the full catalog of planets from the NEA (gray) are shown in Figure~\ref{fig:histMR2}. Histograms on the left include only those planets who have measured values for mass or radius. Histograms on the right include both measured values and calculated values using the mass-radius relationship from \citet{chen2017}. Central values were used when each planet was placed in a bin, uncertainty measurements were not used to determine bin placement for any of the histograms of Figures \ref{fig:histMR2} and \ref{fig:hist2}.

The side-by-side comparison of the distribution of planets with only measured values versus that with a combination of measured and calculated values
highlights a bias in the mass-radius relationship, where massive planets tend to be grouped in the Jupiter radius bin.
There is a smaller fraction of low-mass planets with measured masses which is noticeable when comparing the distribution of planet masses in Figure 2A, which includes only planets with measured masses, to Figure 2B, which also includes planet masses derived from planet radius. This reflects a bias due to the difficulty in measuring the masses of small planets.  
The distribution of planets in the HZ largely follows that of the full catalog of planets, with the Kolmogorov-Smirnov tests performed on Figure 2A indicating that the populations of the full catalog and the 100\% HZ planets came from the same distribution (p-value = 0.608).
There is a slight skew towards larger mass planets of the HZ groups. This is likely due to observational bias, with larger planets being more easily detected further away from their star. 

There is a noticeable gap in the planet mass data in each population, with a peak at $\sim$10~$M_\oplus$ and another at $\sim$400~$M_\oplus$. Similarly, there are two distinct populations of planet radii with a peak at $\sim$2~$R_\oplus$ and another at $\sim$13~$R_\oplus$ in each of the groups of HZ planets and the full catalog of planets. This apparent gap could be attributed to the inherent difficulty of detecting these planets that are typically far out from their host stars, and so rely on detection techniques that are more sensitive to the associated orbital period regime, such as microlensing and direct imaging \citep{Suzuki_2018}. Other theories suggest this dearth of sub-Saturn planets could be attributed to core accretion theory \citep{Ida2004}, where planets that reach 10~$M_\oplus$ enter a runaway accretion period and rapidly grow to $\geq 100$~$M_\oplus$, provided there are sufficient materials available. The existence of this gap is still debated within the literature, such as the work by \citet{Bennett2021} that contests the assertion by \citet{Mayor2011} and \citet{Emsenhuber2021} that the sub-Saturn valley does exist, showing evidence it was missing from their reanalysis of RV planets observed using CORALIE/HARPS. The gap in our data is also present amongst the HZ planets, where the RV signature of a $\geq~10~M_\oplus$ exceeds the limit of current technology for most stars (around a sun-like star this would create a signal of $K\sim1~m/s$, with a larger signal for the same planet around smaller stars), however without a more thorough analysis of the observational biases that affect the detection of these planets we cannot say whether this gap is real.

It should also be noted that the peak of planets at $\sim$2~$R_\oplus$ should not be taken as an indication of the actual peak of smaller radius planets. Many studies have shown that the number of planet occurrence rates increases to smaller sizes and so this peak is more an indication of the limits of current observational tools rather than the true distribution of smaller planets \citep{wittenmyer2011a, kane2016c,fulton2017,hill2018}.

There is a dip in planet occurrence around the small planet radius gap, or Fulton gap \citep{lopez2013,owen2013a,fulton2017} in panels C and D of Figure~\ref{fig:histMR2} at $\sim$1.8~$R_\oplus$ in the full catalog and $\sim$1.1~$R_\oplus$ within the HZ planets. 
The sample of planets used in \citet{fulton2017} were restricted to orbital periods shorter than 100 days, bright stars ($K_p \leq 14.2$), with $T_\mathrm{eff}$ between 4700~K~$< T_\mathrm{eff} <$~6500~K, impact parameters $b \leq 0.7$, and no giant stars. As a result, few HZ planets were included in the \citet{fulton2017} study. However, the Fulton gap may be present in the radius distributions shown in panels C and D of Figure~\ref{fig:histMR2}, both for the full catalog of planets as well as the $>0\%$HZ and $100\%$HZ planetary radii distributions. A recent study by \citet{gupta2019} predicted that the  upper boundary of the super-Earth planet radii sat $\sim$1.1~$R_\oplus$ at 100 days orbital period. Using the scaling law from the paper (Equation 13 of \citet{gupta2019}) this can be extrapolated to estimate a super-Earth planet radii upper boundary of $\sim$1~$R_\oplus$ at 300 days. This prediction matches the gaps found in our data, which is close to $\sim$1.1~$R_\oplus$ in the HZ groups. However, it must be noted that due to small number statistics and the absence of any correction for detection biases and completeness, we are unable to provide a firm conclusion regarding whether the observed gap in the HZ is a reflection of true planetary occurrence rates or observational biases. For example, recent studies of occurrence rates around FGK stars by \citet{Kunimoto2020} and \citet{Hsu2019} found no evidence of the above described radius gap continuing to longer periods. Thus, further study into whether this gap exists in the HZ is essential as it will contribute to the discussion of whether this size bi-modality is caused by photoevaporation, where energy from the host star heats the upper atmospheres of the planet inducing a hydrodynamic outflow, or by core-powered mass-loss, where internal energy from the planet leftover from formation is released into the atmosphere and radiated away causing a hydrodynamic outflow, or by a combination of the two \citep{Rogers2021,Lammer2003, MurrayClay2009,  Ginzburg2018, gupta2019}. In particular, atmospheric loss processes play an important role in the initial conditions and ongoing evolution of terrestrial planet surface conditions \citep{dong2017a}. Given the distance of HZ planets from their star, core-powered mass-loss is the likely contributor to the small planet radius gap within the HZ, if proven to exist. The possibility of significant photoevaporation contributing to this radius gap amongst the HZ planets, however, would have significant implications for the atmospheric evolution of HZ planets and is worth investigating.

It is evident that more of the larger mass planets, and to a lesser extent, more of the larger radius planets, are being excluded from the $100\%$HZ group than smaller mass or radius planets. The planets being excluded from the $100\%$HZ group are also the longer period planets with higher eccentricities (Figure~\ref{fig:hist2}). This hints that the initial origin of these giant planets was beyond the snowline, and migration or planet scattering likely led to their current positioning. 

Histograms of the distribution of planet period, eccentricity, stellar effective temperature, $\log~g$, and $J$ are shown in Figure~\ref{fig:hist2}. In order for liquid water to exist on the surface of a planet, it cannot orbit too close to the star. For that reason the right skew of the peak of the HZ planets period compared to the full catalog was expected. The expected period distribution peak for the HZ planets was in the hundreds of days due to the focus of missions like Kepler on sun-like stars. Considering the star type of each group peaks around G-type stars, the $>0\%$HZ planets peak at $\sim$1000~days orbital period was higher than was initially anticipated. This is due to the inclusion of all planets that spend any amount of time in the HZ; long period eccentric planets that pass through the HZ contribute significantly to skewing the period distribution towards longer periods. For the $100\%$HZ planets, the distribution moved towards smaller periods with a peak of $\sim$200~days. 

As noted earlier, the planets being excluded from the $100\%$HZ group are planets with higher eccentricities, as seen in the eccentricity histogram of Figure~\ref{fig:hist2}. The orbits of planets with high eccentricity will cover a wide range of distances from the star and so are likely to pass through a particular area around the star like the HZ provided their orbital period is long enough. Thus the flat distribution of the $>0\%$HZ planets eccentricity is to be expected. Once the selection was refined to only the 100\%HZ planets it could be seen that the distribution more closely resembles that of the full catalog. As smaller planets have been shown to prefer circular orbits \citep{kane2012d, Van_Eylen_2015}, the low eccentricities of the Rocky-OHZ planets is expected.  

The peak of the TSM distribution for the HZ planets is in the middle of the gap for the full distribution. These low TSM values for HZ planets are likely driven by the $T_\mathrm{eq}$ of these temperate planets.  

Of the remaining histograms in Figure~\ref{fig:hist2}, there is a valley of planets found around K stars ($T_\mathrm{eff} \sim 4000$~K), particularly within the HZ. This is likely due to observation biases; target selection of RV surveys peak around solar type stars as the spectral lines of G-type stars are ideal for RVs. Later type stars have increased activity and earlier type stars lose the spectral lines that the RV method relies on \citep{vanderburg2016}. Stellar rotation periods and associated activity signals can often peak throughout the orbital period range in the HZ of M dwarf stars, creating an additional source of uncertainty for later type stars \citep{vanderburg2016}. Transit studies on the other hand peak at M-dwarf stars as transit surveys are biased towards short period planets due to the necessity of detecting multiple transits for confirmation \citep{Batalha_2010}. Planets transiting M-dwarf stars are also easier to find because of relatively large planet-to-star radius ratios allow greater transit depths for the same planet size.

To demonstrate the bias toward detecting HZ planets around M-dwarf stars with the transit method, we looked at the breakdown of HZ planet detection with the Kepler mission. We adopted a cutoff from \citet{dressing2013} of $T_\mathrm{eff} < 4000~K$ to determine which stars were M-dwarfs within Tables \ref{tab:HZpl} and \ref{tab:HZstar}, with $\log g > 4$ to ensure only main sequence stars were selected. We found 9 of the 86 ($\sim10\%$) Kepler planets in our tables are found to orbit within the HZ of M-dwarf stars, despite the fact that M-dwarf stars constituted only $\sim2\%$ of the Kepler stellar sample \citet{Berger2020}. This bias is even more extreme for the smallest planets, which are near or below the limit of detectability with the transit method. We find $\sim39\%$ of Kepler HZ planets with radii $\leq$2~$R_\oplus$ orbit M-dwarf stars. Bias-corrected occurrence rate estimates from Kepler have shown similar HZ abundances for M and F,G,K dwarf stars \citep{dressing2015b,bryson2021}, indicating that the over-inflation of M-dwarf planets in our Kepler sample is predominantly due to observational biases.

The gap in the $J$ distribution around 8.5-9 magnitude for each population is again due to observational biases. As transit surveys try to simultaneously observe a large number of stars in their field of view, transit host stars tend to be very faint. RV host stars on the other hand tend to be bright, with only a handful facilities performing well on stars fainter than 9th magnitude. The peak of HZ planets around fainter stars is smaller than that of the full catalog. This is an indication that planets in the HZ of dim stars are difficult to find due to the reduction in detectability via some methods like RV and transits with distance from the star.

Figure~\ref{fig:MRscatter} includes mass-radius scatter plots colored with a third parameter: TSM, $J$ or density. Plots A, C and D include only the planets found in the HZ that have both a measured mass and radius. Plot B also includes planets with mass or radius calculated by the method in \citet{chen2017}. The color of data points on the top row of Figure~\ref{fig:MRscatter} indicate TSM of the planet \citep{kempton2018}. On the left, one planet with both measured mass and radius stands out as having a significantly higher TSM value than the other planets: GJ~414~A~b, a sub-Neptune planet that spends 48\% of its orbit in the OHZ initially discovered by \citet{Dedrick2021}. GJ~414~A~b has the greatest TSM of all known transiting planets, including those without mass measurements. This can be attributed to both the radius of the planet and the brightness of the star. On the right in Figure 4B, HD~102365~b, a Neptune-like planet discovered by \citet{tinney2011} that spends 33\% of its orbit within the OHZ, has the highest TSM value of all the planets in the HZ list. Again, the larger planet size and low apparent magnitude of the star contribute to the high TSM value of this planet. It must be noted that HD~102365~b is not known to transit. Many of the planets included in Figure 4B are not known to transit, rather they have been included so that we can directly compare atmospheric characterization of known planets in a broader demographic context.

In plot C the color of data points indicate $J$. The brightest star of the measured values is GJ~414~A. This brightness contributed to the high TSM value of the planet orbiting this star. 

Plot D of Figure~\ref{fig:MRscatter} includes data points colored to the corresponding bulk density of the planet in $g/cm^3$. The planets with the greatest density of all the planets with both mass and radius measurements are the TRAPPIST-1 planets d, e, f and g seen at the bottom left of the plot \citep{gillon2016}. 

Figures~\ref{fig:heat1} and \ref{fig:heat2} include scatter plots of the entire exoplanet catalog (gray) versus the $>0\%$HZ planets (green). If the plot includes mass, radius or eccentricity, only planets with measured values for each are included. These plots also include a heat map to allow easy identification of clusters and their relative density. 
 
Plot A of Figure~\ref{fig:heat1} shows planet period versus planet mass. There are three clusters of planets in the full catalog: Hot Jupiters at the top left, cool Jupiters at the top right and hot super Earths at the bottom left. Of the $>0\%$HZ planets, there is a large cluster of cool Jupiter planets and then a smaller cluster of warm super Earths and terrestrial planets. This plot highlights the relationship between the size of the planet and the stellar type, as the HZ of smaller stars is closer than larger stars and thus the period of HZ planets orbiting M-dwarfs is smaller than those orbiting larger stars.  Within the population of HZ planets shown in the plot, no giant planets appear in the HZ around smaller stars. As giant planets will have a greater gravitational effect on smaller stars, if these planets did exist there is a high likelihood they would have been detected. These results are in line with previous studies that calculated lower occurrence rates of giant planets around low mass stars than higher mass stars due to the lack of planet building materials \citep{montet2014,bowler2015,hill2018}. While the plot also suggests that larger stars tend to host more massive planets in the HZ, this is likely influenced by the increased difficulty encountered when trying to detect lower mass planets around larger stars.   
 
Plot B of Figure~\ref{fig:heat1} shows planet period versus planet radius. There are two major clusters in the full catalog which show the sub-Saturn valley \citep{Ida2004,Suzuki_2018}, but as is mentioned earlier in this section, this gap may be attributed to the inherent difficulty of finding these types of planets, particularly by both the transit and RV detection methods which dominate the detection of exoplanets to date. Similarly to the period vs mass plot, smaller HZ planets tend to be found around smaller stars. 

Plots C, D and E of Figure~\ref{fig:heat1} show eccentricity versus planet orbital period, mass and radius respectively. Longer period and larger mass planets tend to have a wider range of eccentricities than the short period, low mass planets. The closer to the star and the less massive a planet is, the more likely the gravitational pull of the star will force the planet's orbit to circularise.
This is seen to some extent in the radius vs eccentricity plot as well, though less strongly amongst the HZ planets due to the small number of planets with measured radii in the $>0\%$HZ group.

Plot F of Figure~\ref{fig:heat1} shows planet mass versus radius of all the measured planets. The mass radius relation is evident, the HZ planets follow the full catalog of planets along the mass-radius relationship, and two main clusters are present: massive Jupiter planets in the top left and Super Earths in the middle of the plot. 

In the stellar effective temperature vs $\log~g$ plot of Figure~\ref{fig:heat2}, the $>0\%$HZ planets span the same distribution of $T_\mathrm{eff}$ and $\log~g$ as the full catalog, except for amongst the more evolved stars. These stars may not host planets in the HZ, or the properties of the stars may mean detection of any HZ planets is beyond current capabilities. As these stars grow the transit depth of existing planets will reduce making it more difficult to detect transiting exoplanets, and RV surveys tend to focus on main sequence stars rather than highly evolved stars. 


\subsection{Extreme cases in the Habitable Zone}
\label{outliers}

Here we highlight some of the extreme cases in the catalog of HZ planets and discuss how their unique attributes may affect the habitability on these planets. Future observations of these planets will allow testing of these hypotheses and further refinement of the HZ boundaries. 


\subsubsection{Eccentric Planets}
\label{ecc}

Many studies have proposed that planets that are temporarily outside the HZ can maintain or regain habitable periods \citep{williams2002, kane2012e, way2017a, georgakarakos2018, palubski2020a, Kane2021}. Observations of more eccentric HZ planets will help test the boundaries of habitability and the effect of eccentricity on planets orbiting within or passing through the HZ. 

\begin{figure*}
\centering 
\subfloat{%
  \includegraphics[width=0.7\columnwidth, height=0.7\columnwidth]{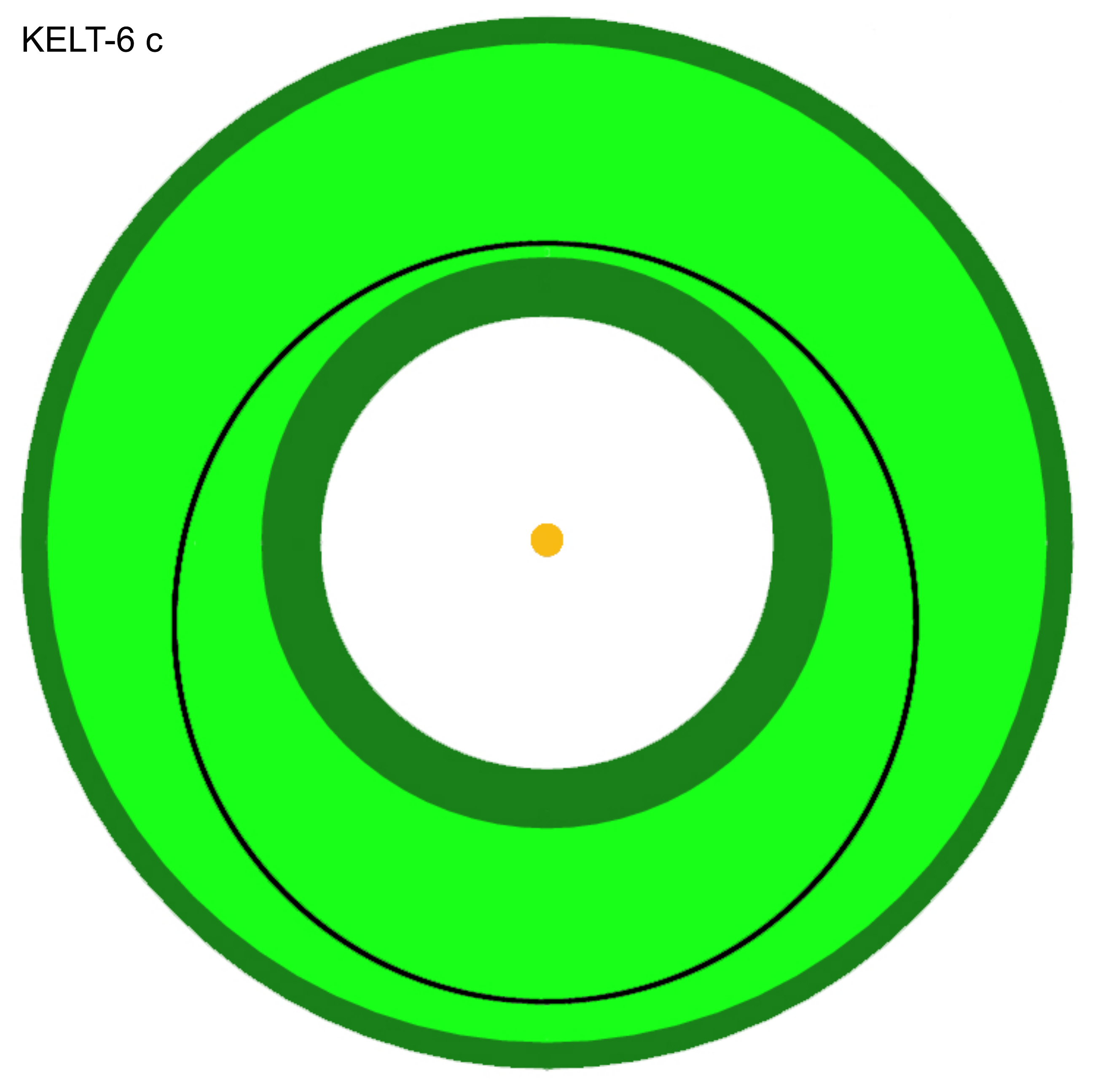}%
}\subfloat{%
  \includegraphics[width=0.7\columnwidth, height=0.7\columnwidth]{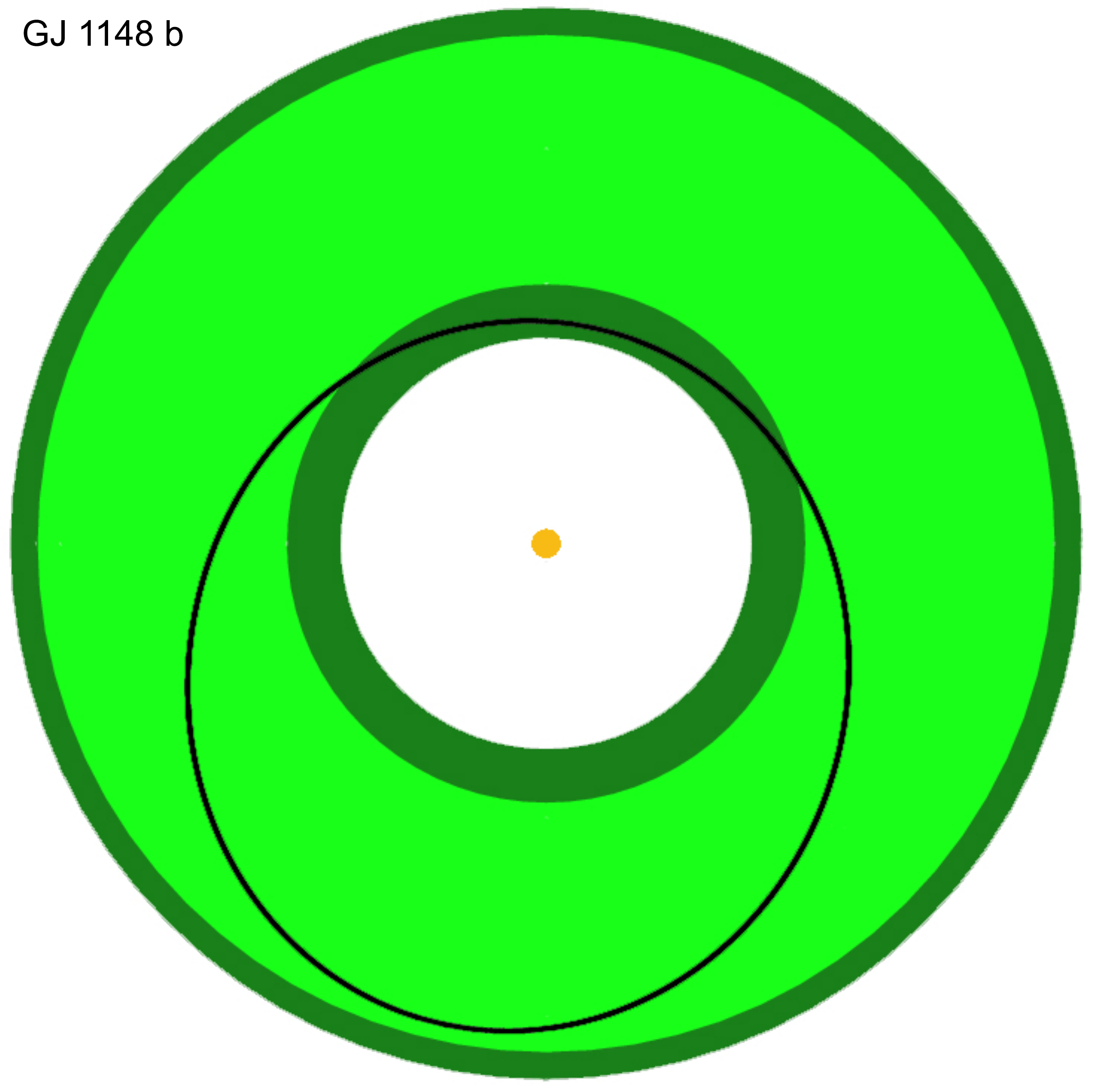}%
}
\subfloat{%
  \includegraphics[width=0.7\columnwidth, height=0.7\columnwidth]{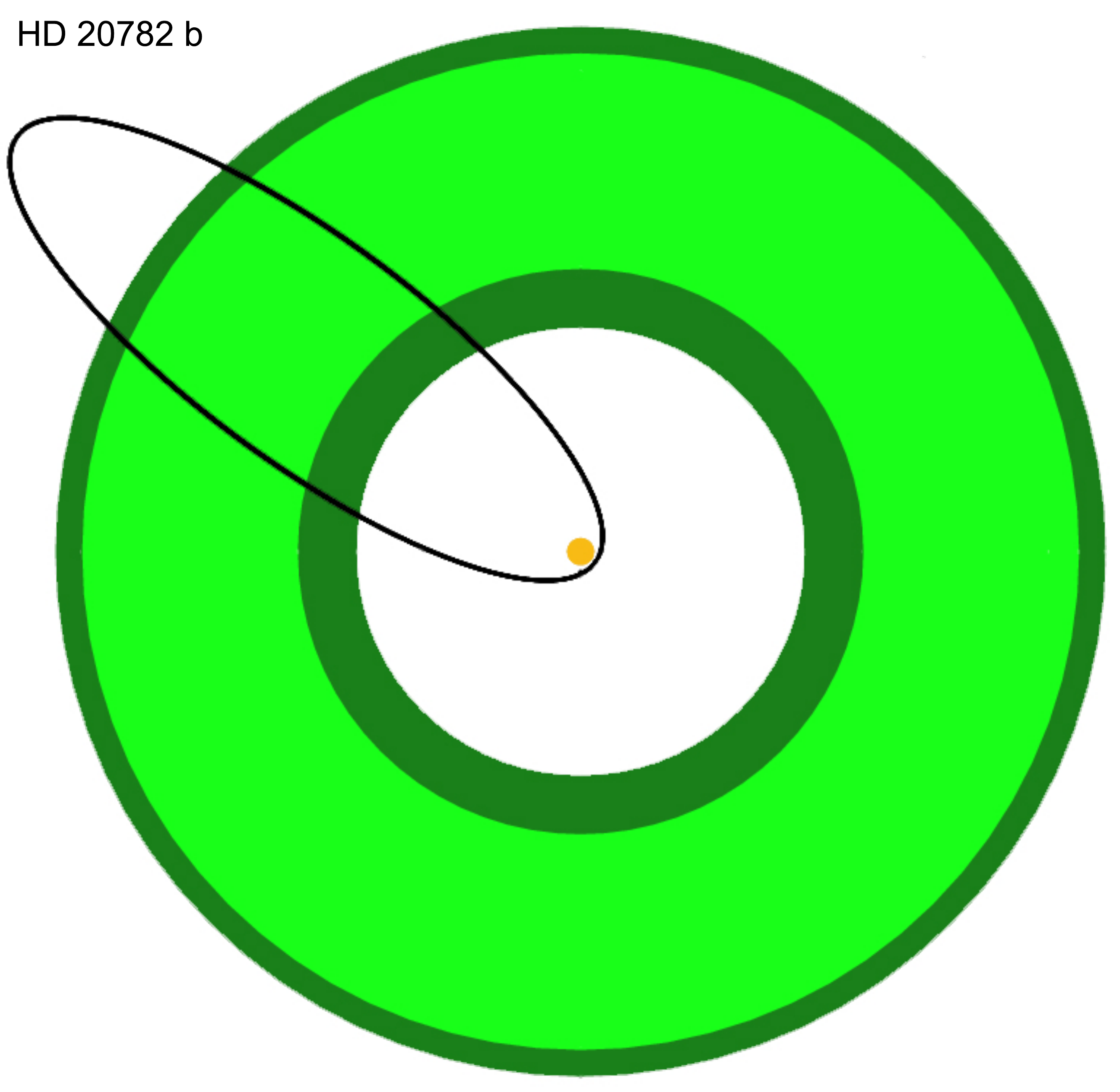}%
}
\quad
\subfloat{%
  \includegraphics[width=0.7\columnwidth, height=0.7\columnwidth]{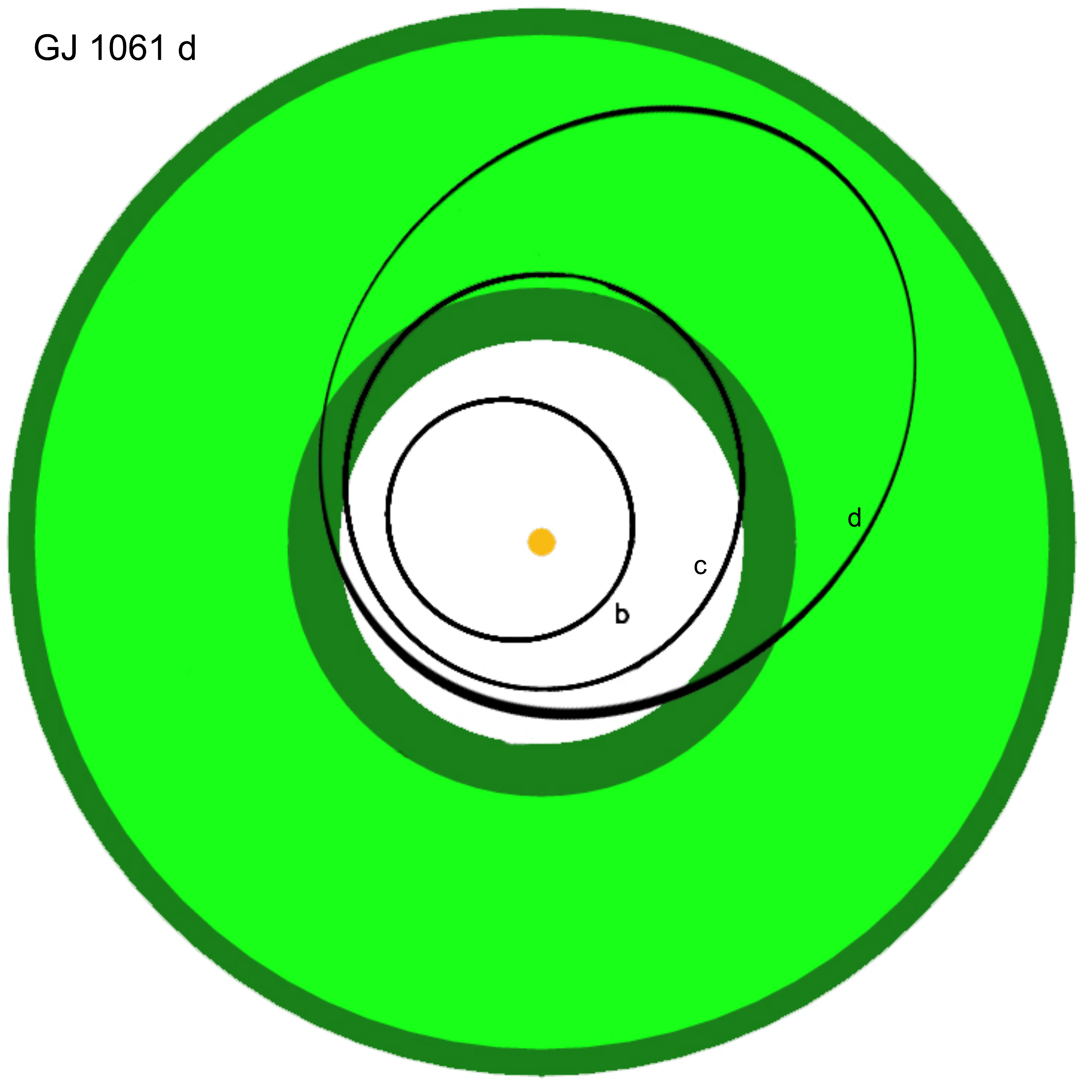}%
}
\subfloat{%
  \includegraphics[width=0.7\columnwidth, height=0.7\columnwidth]{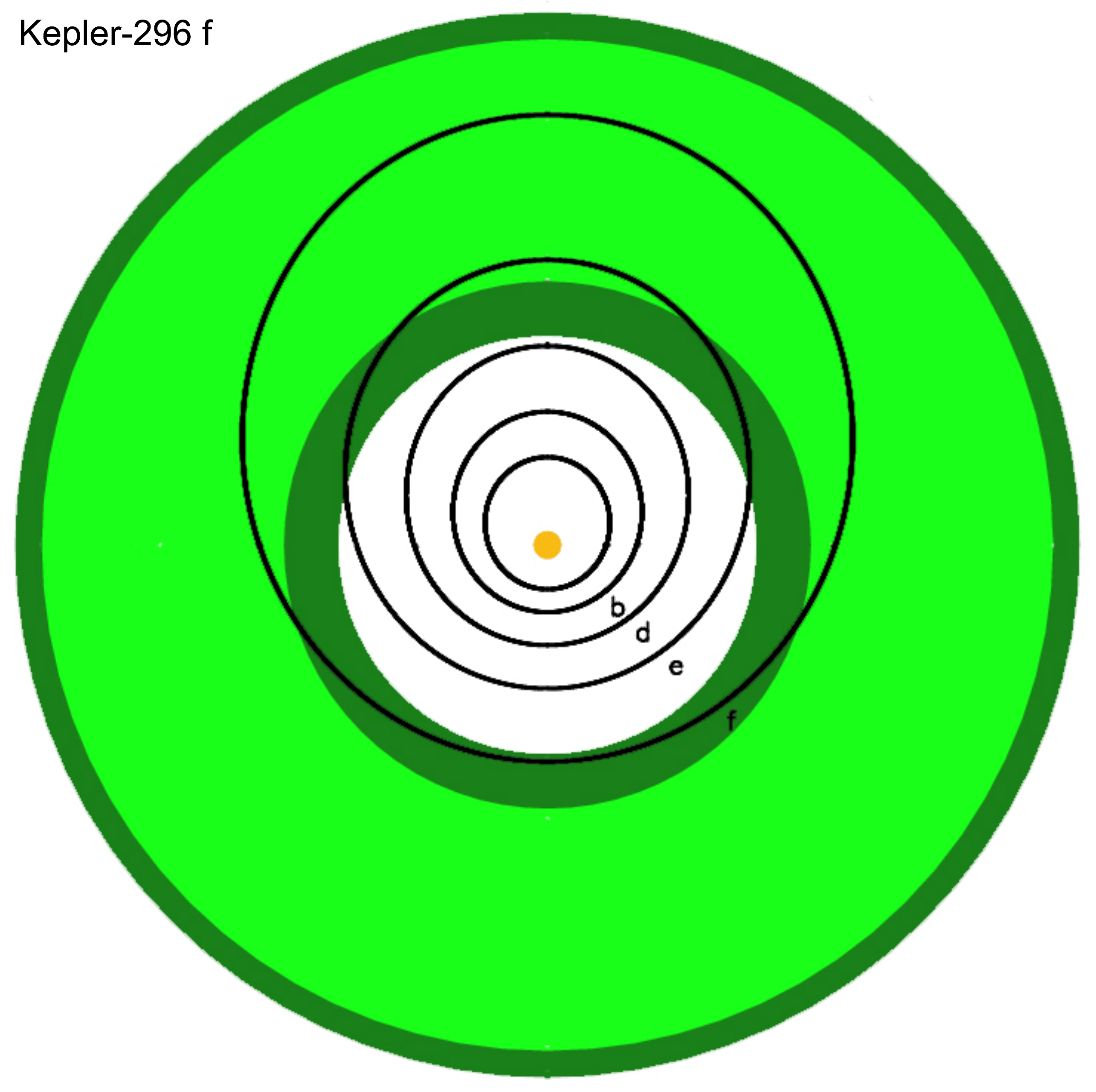}%
  }
\caption{The most eccentric HZ planets. Top left: Most eccentric planet within the CHZ: KELT-6~c is a $3.71\pm0.21 M_J$ planet discovered by \citet{Collins2014}. With an eccentricity of 0.21 this planet's $T_\mathrm{eq}$ will range from 385~K at periastron
to 311~K at apastron. Top middle: Most eccentric planet within the OHZ: GJ~1148~b, discovered by \citet{Haghighipour2010}, is a $0.3043^{+0.0044}_{-0.0032}$~$M_J$ planet. With an eccentricity of 0.38 this planet's $T_\mathrm{eq}$ will range from 422~K at periastron to 283~K at apastron. Top right: Most eccentric planet to pass through the HZ: HD~20782~b, discovered by \citep{jones2006}. This $1.4878^{+0.1045}_{-0.1066}$~$M_J$ planet is the most eccentric planet found to date and has a massive variation in $T_\mathrm{eq}$ of 1603~K at periastron to 257~K at apastron. Bottom left: The most eccentric planet $\leq 2$~$R_\oplus$ that passes through the OHZ, GJ 1061 d has an eccentricity of 0.58 and has a variation in $T_\mathrm{eq}$ of 278 -- 502~K \citep{Dreizler2020}.
Bottom right: The most eccentric planet $\leq 2$~$R_\oplus$ that spends 100\% of its orbit in the OHZ, Kepler-296~f, has an eccentricity of 0.33 and fluctuations in $T_\mathrm{eq}$ of 303--427K \citep{Rowe2014}.
\label{fig:ecc}}
\end{figure*}

The highest eccentricity planets in and passing through the HZ are shown in Figure~\ref{fig:ecc}. The most eccentric planet within the CHZ is KELT-6~c, a $3.71\pm0.21$~$M_J$ planet orbiting 100\% in the CHZ of a F8 star \citep{Collins2014}. With an eccentricity of 0.21, this planet's $T_\mathrm{eq}$ will range from 385~K at periastron to 311~K at apastron. The most eccentric planet within the OHZ is GJ~1148~b, which spends 100\% of the time in the OHZ and 83.9\% in the CHZ. Discovered by \citet{Haghighipour2010}, GJ~1148~b is a $0.3043^{+0.0044}_{-0.0032}$~$M_J$ planet orbiting a M4V star. With an eccentricity of 0.38 this planet's $T_\mathrm{eq}$ will range from 422~K at periastron to 283~K at apastron. 
The most eccentric planet to pass through the HZ is also the most eccentric planet to be found to date: HD~20782~b \citep{jones2006}. This $1.4878^{+0.1045}_{-0.1066}$~$M_J$ planet orbiting a G3V star has an extreme 0.95 eccentricity, causing a massive variation in $T_\mathrm{eq}$ of 1603~K at periastron to 257~K at apastron. 
Kepler-296~f is the most eccentric $\leq 2 R_{\oplus}$ planet that spends 100\% of its orbit in the OHZ of an M2~V star \citep{Rowe2014, muirhead2012}. With an eccentricity of 0.33 this planet undergoes fluctuations in $T_\mathrm{eq}$ between 303--427K.
GJ~1061~d, a $1.64^{+0.24}_{-0.23} M_{\oplus}$ planet, is the most eccentric terrestrial planet that passes through the HZ of an M5.5~V star \citep{Dreizler2020}. With an eccentricity of 0.53, GJ~1061~d spends 87\% of its orbit in the OHZ and the remaining 13\% interior to the zone. The range of $T_\mathrm{eq}$ that GJ~1061~d undergoes is 278-502K.

\setlength{\belowcaptionskip}{2pt}


\subsubsection{Giant Planets}
\label{giant}

The most massive planets ($\leq~13~M_J$) in the HZ are HD~106270~b, a $10.13\pm0.27$~$M_J$ planet orbiting 100\% within the OHZ of a G5IV sub-giant star \citep{Johnson2011} and HD~38529~c, $12.99\pm0.15$~$M_J$ planet that spends 56.8\% in the OHZ of a G8III, slightly evolved star \citep{Fischer2001}. While giant planets are not considered habitable themselves, they may be host to large, terrestrial exomoons. These moons could then benefit from the reflected light and emitted heat from the planet, as well as tidal heating from the motion of the moon around the planet \citep{heller2013a}. With all this on top of the flux received from the host star, a moon orbiting a planet like HD~38529~c, which spends part of its orbit exterior to the HZ, may have an increased chance at maintaining habitable conditions than a lone planet on a similar orbit would \citep{heller2013a, heller2014a, hill2018}. 
Note that we use a conservative upper limit on giant planet mass of $\leq 13$~$M_J$ in this section. This is to ensure no brown dwarfs are included in this section (see Section~\ref{BD} for brown dwarfs in the HZ).


\subsubsection{Brown Dwarfs}
\label{BD}

HD~214823~b is the largest brown dwarf ($20.3\pm2.6$~$M_J$) orbiting entirely within the OHZ of a slightly evolved G0 star \citep{diaz2016}. With 45\% of its orbit within the OHZ, BD-00~4475~b is the largest brown dwarf ($25.05\pm2.23$~$M_J$) passing through the HZ of a G0 star \citep{Dalal2021}. A brown dwarf fusing deuterium will emit heat and thus have its own HZ \citep{Belu2013}. Any satellite orbiting within the HZ of a brown dwarf that is orbiting within the HZ of the its star would potentially have a prolonged period of habitability. Stars increase their luminosity as they age and so during the faint early lifetime of the star, a satellite in the HZ of the brown dwarf may rely on the brown dwarf's emitted heat until the stellar temperature increases sufficiently so that both the satellite and brown dwarf orbit within the HZ of the star. Additionally, a brown dwarf will slowly cool as it ages, but a body orbiting a brown dwarf that orbits within the HZ of the star will be able to maintain temperate conditions long after the brown dwarf has cooled. 


\subsubsection{High and Low Density Planets}
\label{density}

Density was only calculated for planets that had both a measured mass and radius. Of the 13 HZ planets that met these conditions, the highest density planets are TRAPPIST-1 f, g \& e \citep{gillon2016}. Studies into the possible composition of these planets have proposed that these are volatile rich planets \citep{unterborn2018a}. As TRAPPIST-1 is a priority target for observations with the James Webb Space Telescope (JWST), there will soon be more insight into the composition of these worlds.    

The lowest density planets both within and passing through the HZ are also all circumbinary planets. These are Kepler-1661~b, Kepler-1647~b and Kepler-16~b \citep{socia2020, Kostov2016, Doyle2011}. More about circumbinary planets in the HZ can be found in Section~\ref{cb}. 

The lowest density planet that is non circumbinary is GJ~414~A~b, a puffy mini-Neptune that spends 48\% of its orbit in the OHZ of a K7 V star \citep{Dedrick2021}.  

Observations of the atmospheres of the lowest and highest density planets within the HZ will give insight into the composition of these planets and will help in determining the density limitations of habitability.


\subsubsection{Evolved Stars}
\label{evolved}

Numerous planets have been found in the HZ of evolved stars (Figure~\ref{fig:heat2}). While these planets are currently within the HZ of their star, this will be temporary. As stars age they will increase in radius and luminosity and the HZ will move further away from the star. Planets that were once exterior to the HZ will temporarily orbit within the HZ as the star evolves and expands at the end of its life. Many of these planets will have formed exterior to the snow line, the distance from a star where lighter elements will condense \citep{Hayashi1981, MORBIDELLI2015, Lambrechts2014},
and so may have large inventories of $H_2O$ and other volatiles essential for the development of life. It is unclear what length of time is required for life to take hold and evolve on a planet so these planets are excellent targets for observations to determine both if life can start on a planet that spent the majority of its life exterior to the HZ, and if life can evolve to a point where a biosignature can be detected within the time the planet remains in the HZ. Additionally, as giant planets tend to form far away from the star beyond the snow line, gas giants with potentially large exomoons may exist beyond the main sequence HZ of the star. While the moon may have been able to maintain habitable conditions through tidal heating and energy emitted from the host planet, the evolution of the star and movement outwards of the HZ will allow a time in which the moon will exist in the temperate zone of the star. This may enable a period where life that had managed to take hold on the moon can now proliferate, or conversely, may destroy any existing life that is unable to adapt to the changing conditions on the moon. 

There is a noticeable lack of HZ planets detected around highly evolved stars, as seen in Figure~\ref{fig:heat2}. As mentioned in Section \ref{demo}, this is likely due to an observational bias.


\subsection{Circumbinary Planets}
\label{cb}

The HZ boundaries of circumbinary systems has been been a topic of significant interest within the habitability literature \citep{haghighipour2013,kane2013a,cuntz2014,muller2014,cukier2019}. Such interest includes the investigation of the climate effects that the variable insolation flux received by the planet may have upon their climate evolution \citep{mason2015,popp2017,wolf2020}.
There have been numerous CBP discoveries, particularly in the era of Kepler and TESS. The first discovery was that of Kepler-16b, a $0.333\pm0.016$~$M_J$ planet \citep{Doyle2011} whose orbit lies along the inner edge of the HZ \citep{kane2013a}. Other discoveries of CBPs that lie within the binary HZ are Kepler-47~c, a 3.17~$M_\oplus$ planet \citep{orosz2012,orosz2019}, Kepler-453~b, a $\leq~16$~$M_\oplus$ planet \citep{welsh2015}, Kepler-1661~b, a 17~$M_\oplus$ planet \citep{socia2020}, and Kepler-1647~b, a 1.52~$M_J$ planet \citep{Kostov2016}. 

It is interesting to note that each of the three lowest density planets are all circumbinary, however the requirement within this study that the planet have both a measured mass and radius contributes to this relationship. The masses of CBPs can be obtained through the transit timing variations (TTVs) of the binary orbit. As these star systems are eclipsing binaries, the likelihood that the planet will also transit is increased \citep{martin2015a}. This is due to the higher likelihood of alignment between the planet and binary star orbits due to the conservation of angular momentum. This causes an increase in the likelihood of a transit of the planet, which is typically low for planets as far out as the HZ.  

For some of these CBP systems the HZ coincides with the dynamical limit of the system \citep{socia2020,welsh2015}, so the closest a planet can orbit the binary and remain stable happens to be within the HZ. The CBPs and their parameters are shown in Table~\ref{tab:cb}.


\subsection{Controversy in the Habitable Zone}
\label{controversy}

Six planets included in Table \ref{tab:HZpl} are listed as controversial in the NEA. HD~40307~g discovered by \citet{tuomi2013} is refuted by \citet{diaz2016b}, who did not recover a significant signal at the expected period. KIC~5951458~b, discovered by \citet{wang2015}, has subsequently been refuted by \citet{dalba2020b} who suggests stellar multiplicity is the source of this signal. GJ~667~C~e \& GJ~667~C~f were both discovered by \citet{angladaescude2013c} and both suggested to be stellar activity by \citet{robertson2014}. Kepler-452~b, which was discovered by \citet{jenkins2015}, has since been claimed to be instrumental effects by both \citet{mullally2018} and \citet{burke2019}. Also refuted by \citet{burke2019} is
Kepler-186~f, initially discovered by \citet{quintana2014a}. As each of these planets remains in the NEA planet table they have been kept in the table of HZ planets, however we suggest caution should be exercised when following up on any of these planets. 


\subsection{Follow-up Opportunities}
\label{followup}

Follow-up observations of planets in the HZ are essential to determining their habitability potential, assessing the usefulness of target selection tools like HZ boundaries, and testing the habitability limits of other parameters within the HZ. Future missions like the JWST will probe the atmospheres of both terrestrial and gaseous HZ planets and provide insight to their composition. For planets not known to transit, atmospheric characterization may largely rely on emission spectra, whose observations have the advantage of occurring outside of the inferior conjunction window \citep{stevenson2020, mandell2022}. 

The rocky planet ($\leq2~R_{\oplus}$) from Table \ref{tab:HZpl} that spends 100\% of the time in HZ, is known to transit, and has the highest TSM value is LHS~1140~b. Discovered by \citet{dittmann2017b}, and most recently revisited by \citet{lillobox2020}, this rocky super-Earth orbits in the middle of the HZ of a M-dwarf star \citep{kane2018a}. TRAPPIST-1~d has the next greatest TSM of the transiting, rocky, 100\% HZ planets, followed closely by K2-3~d, a super-Earth first detected by \citet{crossfield2015}.

Of all the planets in the HZ with TSM values, including those not known to transit, the rocky planet ($\leq~2~R_{\oplus}$) from Table \ref{tab:HZpl} that spends 100\% of the time in HZ with the highest TSM value is
GJ~667~C~c, a Super Earth discovered by \citet{bonfils2013a}. This is followed closely by Teegarden's~Star~b, a terrestrial planet (1.05~M$_{\oplus}$) that has an orbit within the OHZ \citep{zechmeister2019}.

Each of these systems are ideal targets for future observations as they are multi planet systems with multiple planets orbiting within or nearby the HZ. LHS~1140 hosts another rocky planet interior to the HZ \citep{ment2019}, TRAPPIST-1 is a seven planet system with four planets in the HZ of their star, and K2-3 hosts both a super-Earth and mini-Neptune interior to the HZ. 
GJ~667~C is host to another two Super Earth's that orbit within the CHZ: GJ~667~C~f and GJ~667~C~e \citep{angladaescude2013c} (though this has been disputed, see Section \ref{controversy}), while Teegarden's~Star~c is a terrestrial planet (1.1~M$_{\oplus}$) orbiting within the CHZ. 

Multi planet systems are ideal targets for comparative planetology \citep{weiss2018a}. Similar historical conditions can be assumed for each planet in the system and so how the planets differences in size and distance from their star affect their surface conditions can be seen more directly. Thus comparative planetology can be much more powerful in determining the limits of habitability than comparing many single planets around multiple stars.   

Of the $\leq~2~R_{\oplus}$ planets that pass through the HZ and are known to transit, TOI-2285~b has the highest TSM value \citep{fukui2022}.
Atmospheric observations of this planet will allow insight into how excursions outside of the HZ may affect the habitability of rocky planets.

Of those not known to transit, Wolf~1061~c has the highest TSM value, followed by Ross~508~b and Proxima~Cen~b \citep{wright2016b, astudillo2017, harakawa2022,angladaescude2016}. 
 
Of the HZ planets that have both a measured mass and radius GJ~414~A~b, a $2.57~R_{\oplus}$ planet, has the highest TSM value (Figure \ref{fig:hist2}) \citep{Dedrick2021}. Of all the HZ planets, including those with a calculated mass or radius and whom are not known to transit,  HD~102365~b, a $15.89~M_{\oplus}$ planet, has the greatest TSM, followed by 55~Cnc~f, a $47.77~M_{\oplus}$ planet \citep{tinney2011, fisher2008}. 

Observations of the extreme planets from Section \ref{outliers} are essential to determine the limits of habitability. In particular, observations of the more eccentric terrestrial planets would help determine the limits of temperature variation on habitability. The most eccentric terrestrial planet ($\leq~2~R_\oplus$) that spends 100\% of its orbit in the OHZ is Kepler-296~f \citep{Rowe2014}. Kepler-296 is another multi planet system with an additional eccentric, $\leq~2~R_\oplus$ HZ planet that spends 64\% of its orbit in the OHZ; Kepler-296~e (Figure \ref{fig:ecc}). Observations that can probe the atmospheres of these planets will allow a direct comparison of the effects of eccentricity and the associated temperature changes for planets both within and passing through the HZ. 

Giant planets in the HZ could host rocky exomoons that have the potential to be habitable worlds (Section \ref{giant}). Recent discoveries of possible moon signals indicate we are on the cusp of being able to detect these elusive satellites \citep{kipping2022, Teachey2017}. As well as observations to detect these moons, further RV observations of the most promising HZ giants will aid in refining their orbits. This will enable transit ephemeris refinement to allow for better targeted transit follow-up observations to characterize these planets and search for moons.


\subsubsection{RV observations}
\label{RVFU}

\begin{figure*}
\centering 
\subfloat{%
  \includegraphics[width=1\columnwidth]{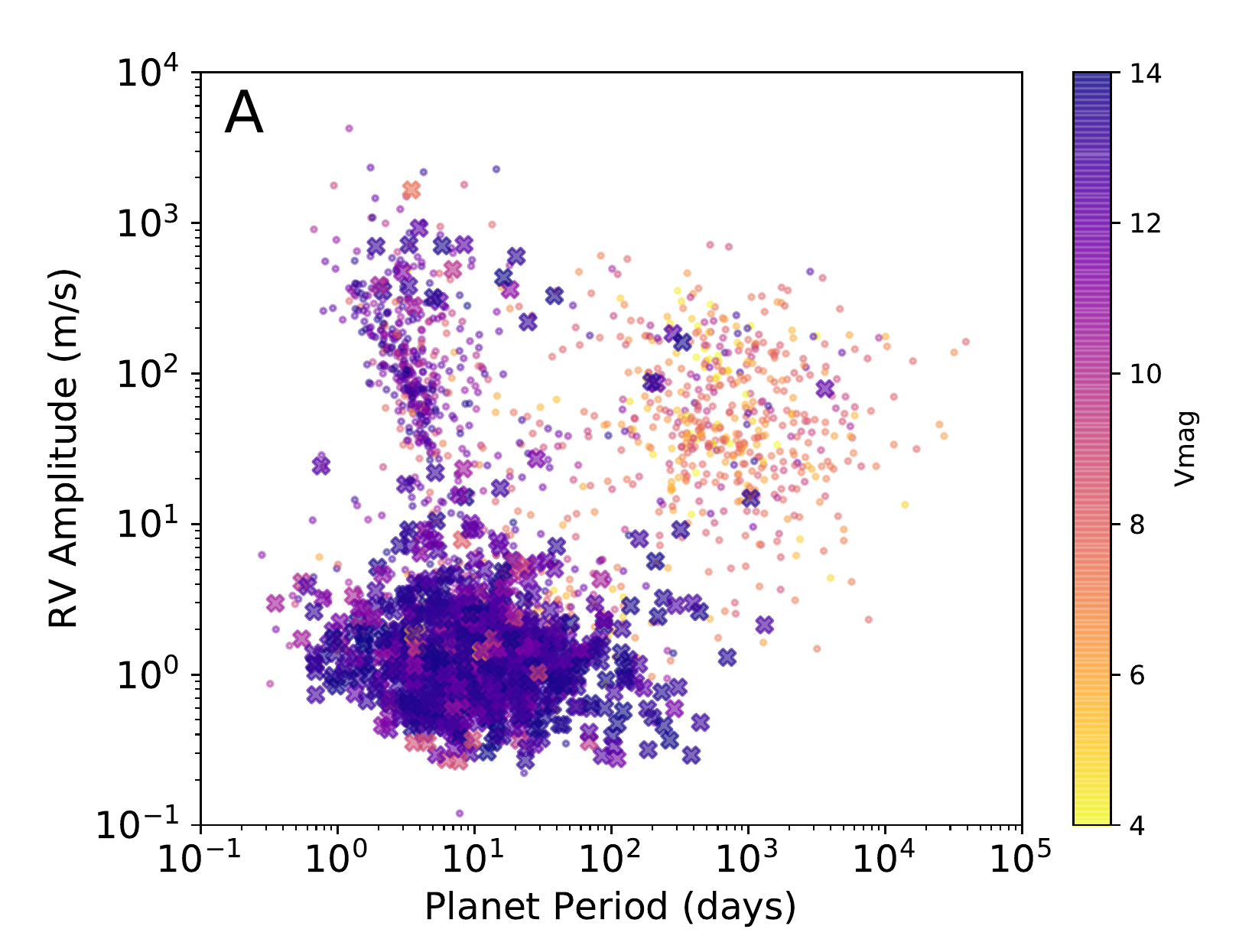}%
}\subfloat{%
  \includegraphics[width=1\columnwidth]{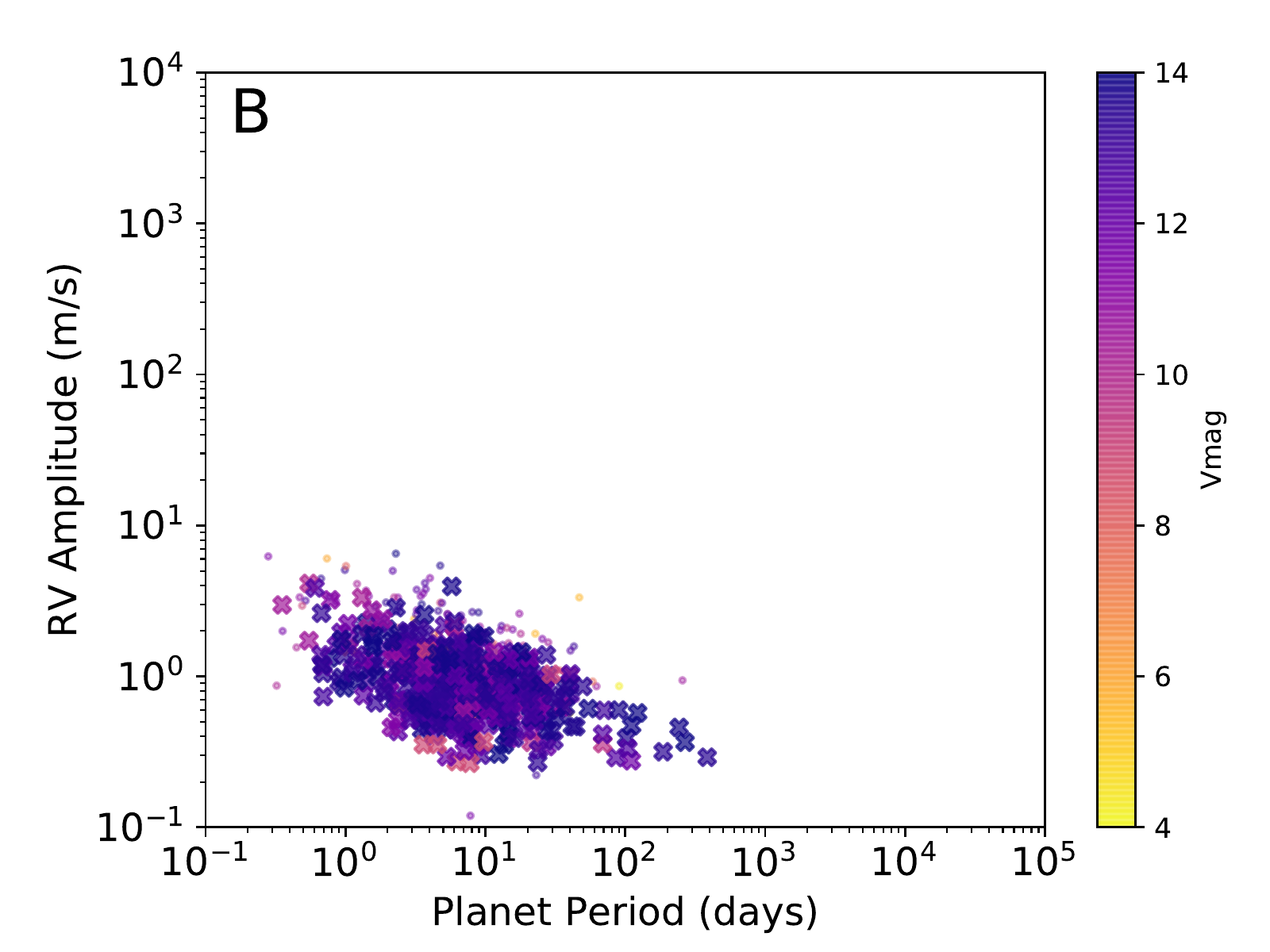}%
}
\quad
\subfloat{%
  \includegraphics[width=1\columnwidth]{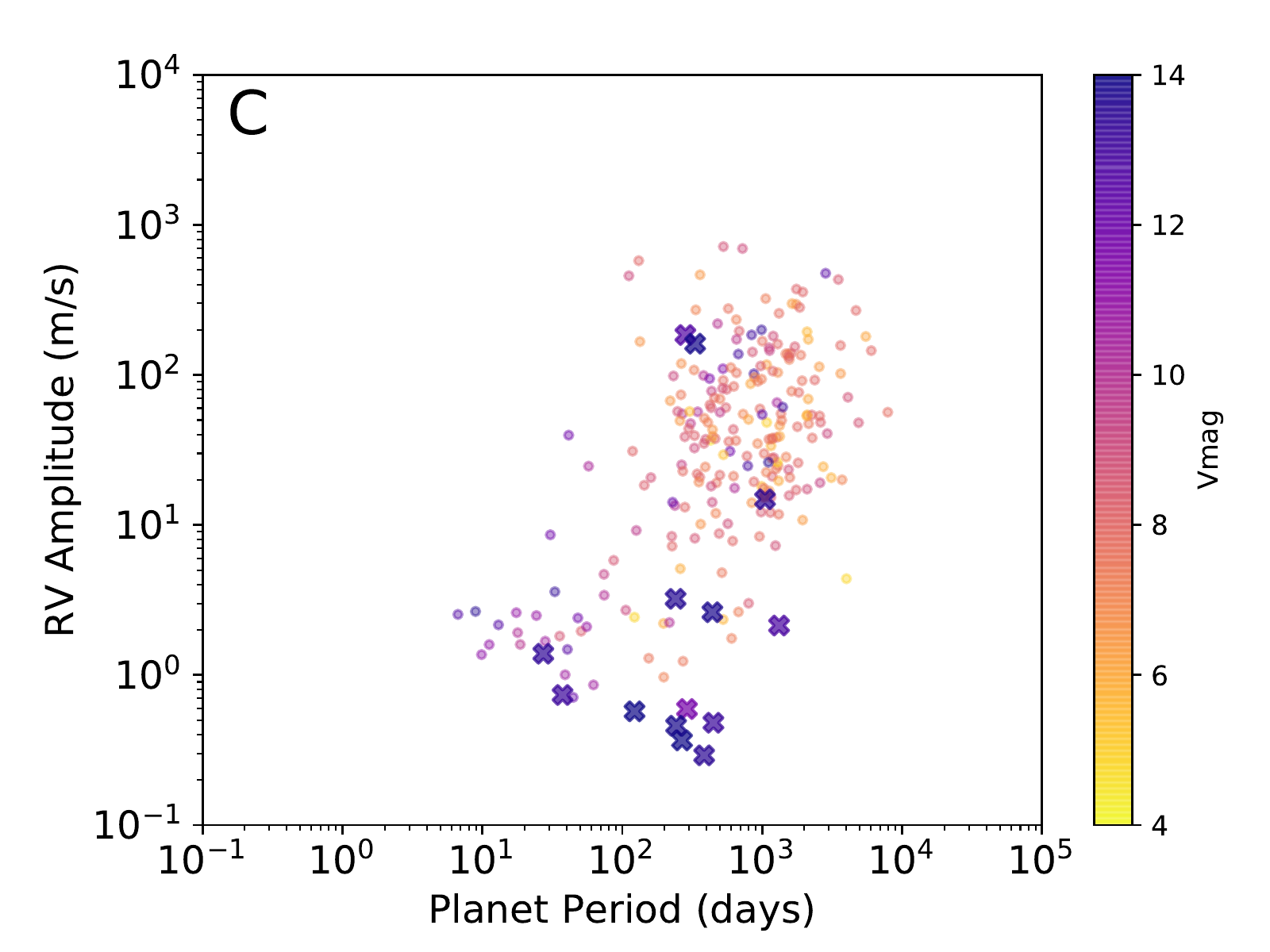}%
}
\subfloat{%
  \includegraphics[width=1\columnwidth]{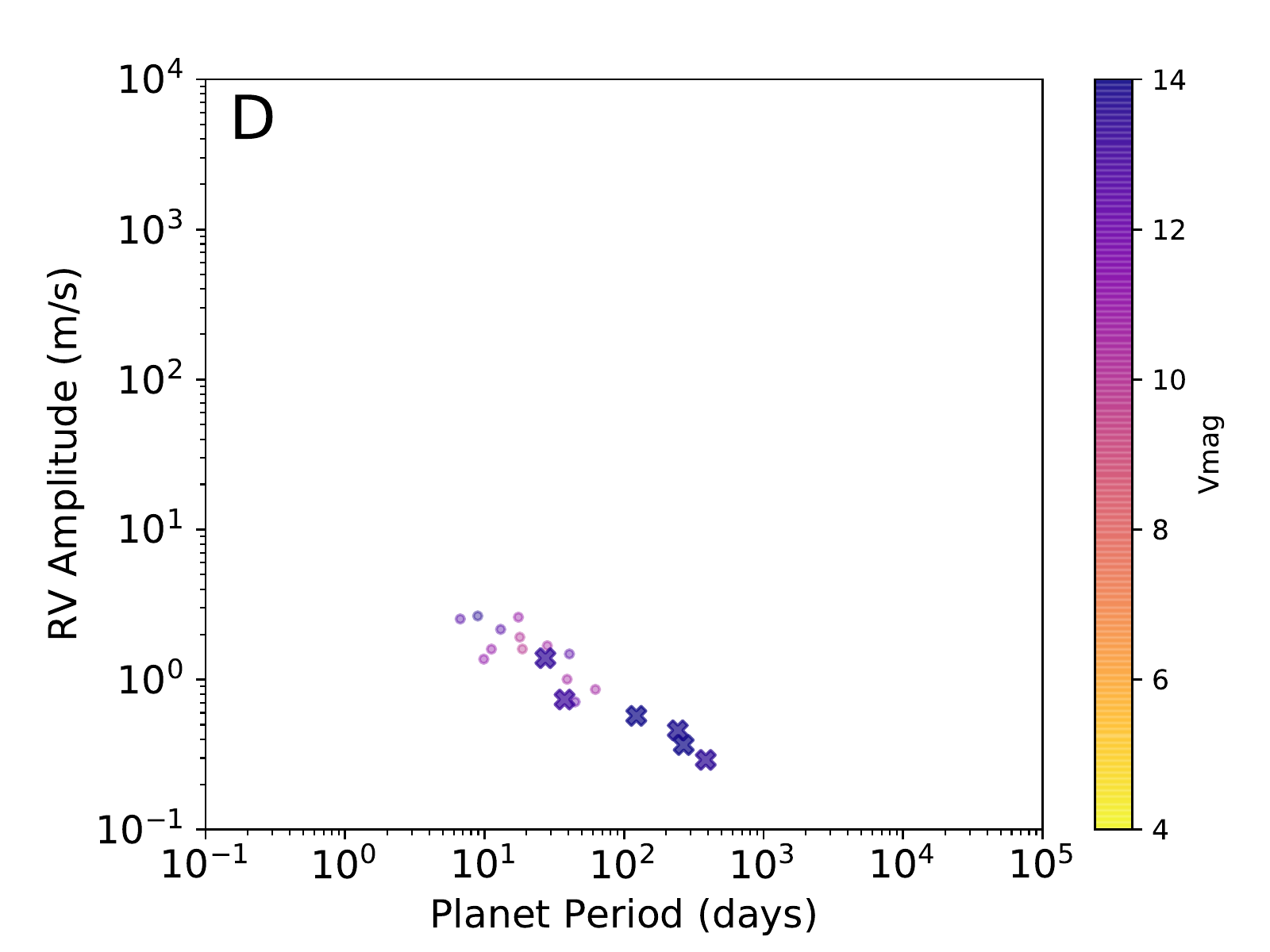}%
  }
\caption{ Predicted RV amplitude as a function of orbital period for four groups of planets: Plot A includes the full catalog of planets, Plot B includes all planets with radii $\leq 2 R_{\oplus}$, Plot C includes the $>0\%$HZ planets and Plot D includes $>0\%$HZ planets with radii $\leq 2 R_{\oplus}$. Planets that are missing mass measurements are denoted by an $X$ symbol. Planets represented with a dot have mass measurements, but many would benefit from RV followup to refine their orbit and mass measurements. The plots are color-coded by $V$ magnitude, as indicated by the color-bar on the right, with a $V$ cutoff of 14 mag. 
\label{fig:RV}}
\end{figure*}

Very few HZ planets have both mass and radius measured. As planet bulk density is an essential measurement to aid in determining atmosphere composition, many more planets with both a measured mass and radius are needed for future transmission missions.
Dedicated RV follow-up of transiting HZ planets and more projects like the Transit Ephemeris Refinement and Monitoring Survey (TERMS) \citep{kane2009c,pilyavsky2011,kane2011b,kane2011c,kane2011e} are needed in order to increase the population of HZ planets that have both and measured mass and radius. 
Figure \ref{fig:RV} shows the predicted RV amplitude as a function of orbital period for four groups of planets: Plot A includes the full catalog of planets, Plot B includes all planets with radii $\leq~2~R_{\oplus}$, Plot C includes the $>0\%$HZ planets and Plot D includes $>0\%$HZ planets with radii $\leq 2 R_{\oplus}$. For planets that were missing mass measurements for the RV amplitude calculations, predicted values from \citet{chen2018a} were used. These planets are denoted by an $X$ symbol, and observations of these planets to obtain mass measurements is paramount. Planets represented with a dot have mass measurements, but many would benefit from RV followup to refine their orbit and mass measurements. 

The plots are color-coded by V magnitude ($V$), as indicated by the color-bar on the right. A magnitude cutoff of $V$ = 14 mag was applied as this is pushing the limit of the best ground-based RV telescopes today. This limit removed the majority of the $>0\%$HZ planets with radii $\leq 2 R_{\oplus}$ (Plot D).

Current state-of-the-art extreme precision radial velocity (EPRV) instruments such as NEID \citep{halverson2016}, ESPRESSO \citep{pepe2021}, and EXPRES \citep{petersburg2020} are capable of achieving sub 30 cm/s single measurement precision (SMP) and around 50 cm/s long-term precision over several months for nearby bright stars \citep{pepe2021}. Such instruments would significantly improve the mass measurements for planets around nearby stars that already have mass values measured thanks to their improved RV precisions over their predecessors, which typically is on the order of a few m/s. Spectrograph performance, however, is highly dependent on the signal-to-noise ratio (SNR) of each measurement and RV measurement precision generally decreases for fainter stars because of the difficulty of achieving high enough SNR \citep{fischer2016}. Obtaining robust mass measurements for HZ terrestrial planets around faint stars that induce RV semi-amplitudes of only $<~50$ cm/s may still be out of the current capabilities even with the latest EPRV spectrographs.

In order to increase the number of planets in this group, investment into ground-based large aperture telescopes and EPRV instruments are paramount to the complete characterization of HZ terrestrial planets and their atmospheres.

\setlength{\belowcaptionskip}{2pt}


\subsection{Additional Notes}
\label{notes}

Only the default values were downloaded from the NEA for each planet included in this paper. The NEA designates a single paper per planet from which it extracts the default planet and star parameters. Because of this, there are instances where the default parameters for a planet is taken from a paper that does not provide measurements for all parameters, even though there are other papers that have measured the parameter (i.e. a RV paper that does not state the radius measured in a previous transit survey). As mixing parameters from different studies can lead to errors due to inconsistent methodology, we elected to only use the default parameters and calculate missing parameters where possible. 
We recommend the reader investigate all papers listed on the NEA for any particular planet of interest to determine the best parameters available. 


\section{Conclusions}
\label{conclusions}



In this paper, we present a complete catalog (at time of writing) of planets that orbit within or through the HZ, including the HZ boundaries and the percentage of the orbital period that each planet spends in their star's HZ. Observational metrics for each planet, such as TSM values and RV amplitude, are included to facilitate selection for future follow-up observations. 
Demographics of Rocky-OHZ, 100\%HZ, $>0\%$HZ, and the full catalog of exoplanets are explored in Section~\ref{discussion}, and various planets are highlighted as potential targets to test the boundaries of habitability.

Histograms of Rocky-OHZ, 100\%HZ, $>0\%$HZ, and the full catalog of exoplanets are discussed in Section~\ref{discussion}. There is a lack of planets in the sub-Saturn radius valley for the 100\%HZ, $>0\%$HZ, and the full catalog of exoplanets. Observational biases may be driving this gap but as the gap is seen amongst the HZ planets, where the RV signature of a $\geq~10~M\oplus$ is well within the capability of detection by current technology for most stars, we believe further investigation into whether this gap is real is warranted.  

The small planet radius gap may be evident in each of the 100\%HZ, $>0\%$HZ, and full catalog groups of planets. Due to small number statistics driven by the difficulty in finding low mass planets in the HZ we are unable to conclude whether this gap is real or due to observational biases. If the gap does exist, determining whether this gap amongst the HZ planets is caused by photoevaporation, core-powered mass-loss or some combination of the two will provide insight into the early atmospheric evolution of these HZ worlds. Each atmosphere loss mechanism has different implications for the atmospheric evolution and surface conditions of a planet, particularly of terrestrial HZ planets. Determining the existence and cause of this gap within the HZ planets is paramount.

Rocky-OHZ planets have predominantly low eccentricities. In general, longer period and larger mass planets tend to have a wider range of eccentricities than the shorter period planets.

TSM values for HZ planets are relatively low compared with the full inventory of known exoplanets, mostly due to the low $T_\mathrm{eq}$ of these HZ planets.
The planets with the highest TSM values that are $\leq~2~R\oplus$, orbit within the HZ and are known to transit are LHS~1140~b, TRAPPIST-1~d and K2-3~d. Of the $\leq~2~R\oplus$, 100\% HZ planets not known to transit GJ~667~C~c and Teegarden’s~Star~b have the highest TSM values, while Wolf~1061~c, Ross~508~b, and Proxima~Cen~b have the highest values for $\leq~2~R\oplus$ planets that pass through the HZ.
Of all the planets ($\leq~10~R\oplus$) passing through and orbiting within the HZ, including those that are not known to transit, HD~102365~b and 55~Cnc~f have the greatest TSM values.  

Observational biases drive many of the features seen in the demographics. Transit surveys are optimised to discover planets around smaller, fainter stars and RV surveys are optimised for brighter, sun-like stars. This causes a bi-modal distribution and a possibly misleading valley to appear in the histograms of both $J$ and $T_\mathrm{eff}$. 
In the planet period versus planet mass plot of Figure \ref{fig:heat1}, the plot suggests that larger stars tend to host more massive planets in the HZ, however this is likely influenced by the increased difficulty in trying to detect lower mass planets around larger stars. 
The lack of HZ planets detected around highly evolved stars, as seen in Figure~\ref{fig:heat2}, is again likely due to an observational bias. Detecting HZ planets around these massive stars may be beyond current capabilities, as the transit depth of existing planets reduces as these stars expand, and RV surveys tend to focus on the more stable main sequence stars.

The inventory of HZ planets will continue to increase as current exoplanet surveys, such as TESS \citep{ricker2015} and CHEOPS \citep{broeg2014}, provide new discoveries and characterization opportunities. With the recommendations of the Decadal Survey on Astronomy and Astrophysics 2020 focusing on the detection and characterisation of habitable exoplanets \citep{Decadal2021}, observations of known HZ planets from this catalog, particularly of the more extreme planets such as those described in Section~\ref{outliers}, are vital. These observations will help determine if the HZ is a useful tool and will facilitate testing of the boundaries of habitability. These more extreme planets are important targets for future observations as they will provide insight into the habitability of planets with high eccentricity, high or low density, circumbinary planets or those orbiting evolved stars along with positioning in and around the HZ. It is essential to follow-up many of these edge case planets to thoroughly test the boundaries of habitability and the robustness of planetary atmospheric evolution. The JWST is set to observe some of the highest measured density HZ planets, such as the TRAPPIST-1 system \citep{lustigyaeger2019a}. Detection and characterisation of any atmospheres on TRAPPIST-1 d,e,f, and g will provide useful insight into the habitability of planets around M-dwarf stars, as well as allow comparative planetology of multiple planets within the HZ. As a similar historical stellar environment can be assumed for each planet, this direct comparison of terrestrial planets at varying distances within the HZ is ideal for determining how habitability changes with position within the HZ. 

The list of known exoplanets continues to grow, with over 5000 confirmed planets and many more awaiting confirmation. While large pools of planets are particularly good for statistical and demographic studies, it presents a significant target selection challenge in the search for potentially habitable planets in quantifying which may be most suitable from both an astrobiological and observation perspective. This is where target selection tools such as the HZ are most useful. By refining the full catalog of planets to only those whose orbits lie within an area around the star where conditions are most conducive to the existence of liquid water on the surface, limited telescope time can be directed to the most promising targets. Observations of the planets in this catalog with JWST and future missions, such as the Nancy Grace Roman Space Telescope \citep{Roman2021}, LUVOIR \citep{reportluvoir} and HabEx \citep{reporthabex}, will provide further insight into the habitability of exoplanets found in the HZ. By using this catalog and the planets highlighted within it to guide target selection, the boundaries of habitability can be investigated and the reliability of the HZ hypothesis assessed.


\section*{Acknowledgements}

M.H. would like to acknowledge NASA support via the FINESST Planetary Science Division, NASA award number 80NSSC21K1536. M.H. Would also like to thank Hilke Schlichting for her advice regarding the radius valley of HZ planets. S.R.K. and C.O. acknowledge support from NASA grant 80NSSC21K1797, funded through the NASA Habitable Worlds Program. T.F. acknowledges support from the University of California President's Postdoctoral Fellowship Program. This research has made use of the NASA Exoplanet Archive. We obtain the data set from the NASA Exoplanet Archive \citep{ps}\footnote{Accessed on 2022-05-19 at 12:45, returning 4550 rows.}. This dataset or service is made available by the NASA Exoplanet Science Institute at IPAC, which is operated by the California Institute of Technology under contract with the National Aeronautics and Space Administration.
This research has also made use of the Habitable Zone Gallery at hzgallery.org.



\end{document}